\documentclass[a4paper,12pt]{article}
\usepackage{amsmath}
\usepackage{bm, bbm}
\usepackage{amssymb,amsthm,graphicx}
\usepackage[titletoc,title]{appendix}
\usepackage{enumitem}
\usepackage{color}
\usepackage{epsfig}
\usepackage{graphics}
\usepackage{pdfpages}
\usepackage{subcaption}
\usepackage[font=small]{caption}
\usepackage[hang,flushmargin]{footmisc} 
\usepackage{float}
\usepackage{booktabs}
\usepackage[mathscr]{euscript}
\usepackage{natbib}
\usepackage{setspace}
\usepackage{mathrsfs}
\usepackage{mathtools}
\usepackage[left=2.7cm,right=2.7cm,bottom=2.7cm,top=2.7cm]{geometry}
\definecolor{royalazure}{rgb}{0.0, 0.2, 0.65}
\definecolor{amaranth}{rgb}{0.9, 0.17, 0.31}

\usepackage{multibib}
\newcites{apndx}{References in Appendix}
\usepackage[pdftex,hypertexnames=false,linktocpage=true]{hyperref}

\usepackage{enumitem}

\newcommand{\reals}{\mathbb{R}}
\newcommand{\integers}{\mathbb{Z}}
\newcommand{\naturals}{\mathbb{N}}

\newcommand{\pr}{\mathbb{P}}        % probability
\newcommand{\ex}{\mathbb{E}}        % expectation
\newcommand{\var}{\textnormal{Var}} % variance
\newcommand{\cov}{\textnormal{Cov}} % covariance

\newcommand{\law}{\mathcal{L}} % law of X
\newcommand{\bfbeta}{\bm{\beta}} %bold beta coefficent
\newcommand{\bfvareps}{\bm{\varepsilon}} %bold varepsilon - vector of standardized residuals
\newcommand{\X}{\boldsymbol{X}} %bold X - vector of covariates
\newcommand{\F}{\mathcal{F}} %sigma-algebra
\newcommand{\grid}{\mathcal{G}_T} %grid
\theoremstyle{plain}

\newtheorem{theorem}{Theorem}[section]
\newtheorem{prop}[theorem]{Proposition}

\newtheorem{remark}[theorem]{Remark}

\newtheorem{definitionA}{Definition}[section]

\newtheorem{lemmaA}[definitionA]{Lemma}

\makeatletter
\newcommand{\lefteqno}{\let\veqno\@@leqno}
\makeatother

\newcommand{\Z}{\boldsymbol{Z}} %bold Z
\newcommand{\x}{\boldsymbol{x}} %bold x
\newcommand{\veca}{\boldsymbol{a}} %bold a
\newcommand{\vecb}{\boldsymbol{b}} %bold b
\newcommand{\vecc}{\boldsymbol{c}} %bold c
\newcommand{\vv}{\boldsymbol{v}} %bold v
\newcommand{\zeros}{\boldsymbol{0}} %bold 0
\newcommand{\bfgamma}{\bm{\gamma}} %bold gamma
\newcommand{\indic}[1]{\mathbbm{1}{\left\{#1\right\}}}

\title{Multiscale Comparison\\ of Nonparametric Trending Coefficients}
\author{Marina Khismatullina\thanks{E-mail: khismatullina@ese.eur.nl. Erasmus School of Economics, Erasmus University Rotterdam} \and Bernhard van der Sluis\thanks{E-mail: vandersluis@ese.eur.nl. Erasmus School of Economics, Erasmus University Rotterdam}}

\begin{document}
\renewcommand{\baselinestretch}{1.2}\normalsize
\maketitle

\begin{abstract}
\noindent
      This paper proposes a novel framework to test for slope heterogeneity between time-varying coefficients in panel data models. Our test not only allows us to detect whether the coefficient functions are the same across all units or not, but also determines which of them are different and where these differences are located. We establish the asymptotic validity of our multiscale test. As an extension of the proposed procedure, we show how to use the results to uncover latent group structures in the model. %Moreover, we conduct a simulation study that showcases good finite sample performance of our method, particularly correct size and satisfactory power for various specifications of the errors and regressors.
      We apply our methods to test for heterogeneity in the effect of U.S. monetary shocks on $49$ foreign economies and itself. We find evidence that such heterogeneity indeed exists and we discuss the clustering results for two groups.


\end{abstract}

\newpage
\section{Introduction}
% heterogeneity in panel data
The use of panel data is a fundamental tool in empirical research across various disciplines, including economics, finance, and climatology. The double-index structure of panel data allows researchers to capture both cross-sectional heterogeneity and time dynamics, making it indispensable for analyzing complex systems. However, the majority of panel data models assume time-invariant and/or common coefficients, assumptions that are often unrealistic in practical applications. Structural changes, policy interventions, and macroeconomic shocks frequently induce time variation in coefficients, and failure to capture such variation can lead to misleading conclusions. Similarly, cross-sectional units may respond differently to covariates. For instance, in finance, the sensitivity of asset returns to market return is often assumed to fluctuate over time \citep{Chen1982, Jagannathan1996, Lewellen2006}. Moreover, investors are interested in the different risk exposures across assets for the same predictors \citep{Fama1973, Adrian2009}. In environmental research, the impact of climatic factors varies across regions and time \citep{Lobell2014, Keane2020}. These examples highlight the importance of developing statistical tools to analyze heterogeneous time-varying coefficients in panel data.

%research question
In this paper, we propose a new multiscale test that allows us to assess heterogeneity in time-varying regression coefficients in panel data models. Unlike traditional global hypothesis tests, which provide no insight into localized differences, our method offers a more detailed analysis by detecting both the presence of coefficient heterogeneity and its specific locations and time intervals. This level of detail makes the test particularly useful in practical applications, where understanding the specific structure of heterogeneity is crucial for policy making, risk management, and forecasting.

% contributions
% 1. informative test for homogeneity: which and where
While several methods have been developed to allow for time-varying coefficients in panel data, they typically focus on global tests that do not identify where or when differences arise. Currently there exists limited theoretical guidance on (i) testing whether time-varying coefficients are all the same across units, and (ii) identifying specific time periods and units where coefficients differ. Existing methods, such as those proposed by \cite{Atak2025}, allow researchers to test for global slope homogeneity, but do not provide localized information where these differences occur. However, simply detecting heterogeneity is often insufficient: practitioners need to understand which time periods and which groups of units drive these differences. Our proposed approach addresses this gap directly.

%Despite growing interest in time-varying coefficient models, existing methods remain limited in several respects. In particular,   In many applications, however, simply knowing that heterogeneity exists is not sufficient: practitioners also need to understand which time periods and which groups of units drive these differences.

% 2. theoretical contributions
The main theoretical contribution of this paper is establishing that under mild regularity conditions, the proposed multiscale test accounts for the multiple testing problem inherent in this setting by controlling the family-wise error rate. Building on the strong approximation results for random vectors developed in \cite{Karmakar2020}, this result ensures the validity of simultaneous confidence statements about localized differences in trending coefficients. This extends previous work on the comparison of nonparametric time trends by \cite{KhismatullinaVogt2022, KhismatullinaVogt2023} to the time-varying coefficient models.

% % linear restrictions; clustering
% We further extend our framework in two important ways. First, we show how our method can be used to test linear restrictions on time-varying coefficients. This allows practitioners to assess hypotheses about, for example, proportional responses across units. \commentsB{better example + sources needed.} Second, following \cite{KhismatullinaVogt2022}, we propose a clustering algorithm that identifies groups of units exhibiting similar time trends. Unlike existing clustering approaches that rely on structural breaks or latent structures, our method is based on a formal statistical test, ensuring that identified groups come with confidence guarantees.

% LITERATURE REVIEW
This paper contributes to several strands of literature. First, the assumption of homogeneity of slope coefficients, whether they are assumed to be time-varying or not, is crucial for valid inference. If the coefficients are truly homogeneous and the individual heterogeneity is captured by fixed effects, cross-sectional data can be pooled to estimate coefficients more efficiently. However, if heterogeneity exists in the coefficients, cross-sectional averaging results in misleading inference \citep{Hsiao1997, Baltagi2008, Wang2019}. Several methods have been developed to address this issue, both for testing for homogeneity in panel data models with constant slope \citep{Pesaran2008, Gao2020, Su2013, Ando2015} and with time-varying coefficients \citep{Atak2025}. However, as mentioned before, these approaches were developed for testing a global null hypothesis that states that the coefficients are the same for \emph{all} units at \emph{all} times, whereas our method provides additional information in case of slope heterogeneity. Specifically, by considering a multitude of localized null hypotheses instead of one global null, we are able to say when and where this heterogeneity exactly occurs.
% develop a test for homogeneity in linear panel data models with constant slope coefficients. \cite{Gao2020} propose a test for homogeneity of constant slope while allowing for heterogeneous nonparametric time trends. \cite{Su2013} and \cite{Ando2015} develop tests for constant slope homogeneity in a large-dimensional panel data models with interactive fixed effects. Recently, \cite{Atak2023} have developed a test for homogeneity and/or stability of time-varying coefficients. Nevertheless, the tests above all consider a global hypothesis: the coefficients are homogeneous for \emph{all} units. We break this hypothesis up into a multitude of local hypotheses that allows us to test at what time and for which units the coefficients are homogeneous simultaneously. We refer to such a test as a multiscale test.

Second, we extend prior work on multiscale testing to a panel data setting with time-trending coefficients. Multiscale tests were developed for various models and null hypotheses. For example, \cite{KhismatullinaVogt2020} propose a new multiscale approach to test for monotonicity of deterministic trend functions in a single nonparametric regression. This framework was later extended to multiple time series by \cite{KhismatullinaVogt2022} and \cite{KhismatullinaVogt2023}, where the authors devise procedures to test for homogeneity of trend functions under different assumptions. In this paper, we build upon the theoretical work in the above studies to consider time-varying coefficient functions. Similarly as before, the objects of interest are functions of time that are estimated nonparametrically and compared on a wide range of scales, i.e., using various bandwidths. This is the essence of a multiscale approach, and this approach allows us to provide information on top of simply rejecting the null hypothesis of a global slope homogeneity.

%There exists a vast literature for estimating nonparametric time-varying coefficients \citep[e.g., ][]{robinson1991time, Cai2007, li2011non}. We use the local constant approach \citep{robinson1991time}, avoiding the first-order bias in the local linear approach \citep[among others]{Cai2007, friedrich2024sieve}.

% The current paper builds upon several important works. First, our test is justified under a strong approximation for random vectors, recently established by \cite{Karmakar2020}. \commentsB{Other relevant, previously established results: \cite{Mies2023, ZhouWu2010, Xu2022, Wu2011,Karmakar2022,Cattaneo2022, Li2020}.} Second, we extend previous tests for slope homogeneity presented by \cite{Chen2019, Chen2021, Pesaran2008, Su2013, Atak2023}. These works have at least one of the two restrictions: (i) time-invariant coefficients; (ii) no local information on the test. We relax these restrictions.  

% In this paper, we assume that the predictors are common across units. Examples of models with common predictors include factor models and many empirical models (), making our test suitable for empirical research. This assumption allows us to rescale the test statistic with the variance of the common predictors, resulting in a Wald-type test that is easy to implement. We leave the extension to unit-specific regressors for further research. 

Furthermore, in our paper, we propose a clustering algorithm that identifies groups of units exhibiting similar time trends. Specifically, in case of detected heterogeneity among the coefficient functions, it is still possible to allow for grouped patterns of heterogeneity that can be a result of similar underlying characteristics across units. %\cite{Bonhomme2022} argue that a group specification can be an accurate discrete approximation of the unobserved heterogeneity, whereas \cite{hahn2010panel} provide foundations for grouped heterogeneity from a game-theoretic framework.
Several approaches exist to identify the unobserved group structure. In the seminal paper of \cite{BonhommeManresa2015}, a $k$-means algorithm is used to cluster the individual observations in a linear panel data model with group-specific fixed effects. This framework has been extended to time-varying slope coefficients, but the extensions are mostly limited to structural breaks \citep{Okui2021, Lumsdaine2023}. Alternatively, \cite{Su2016} propose a classifier-Lasso simultaneous estimator of heterogeneous coefficients and group membership in linear and non-linear panel data models and \cite{Su2019} use the classifier-Lasso in a panel data model with group-specific time-varying coefficients. Their approach requires a sieve approximation of the time-varying coefficients together with a penalized least-squares objective function. Recently, this work has been extended by \cite{Haimerl2025} who instead of the classifier-Lasso use a pairwise adaptive group fused-Lasso to uncover a hidden group structure. In contrast, our estimator of the time-varying coefficients has a closed-form solution, the multiscale test statistics are easy to compute, and the clustering procedure is a straightforward application of an easy-to-track hierarchical agglomerative clustering algorithm. The most similar approach to ours was taken in \cite{Chen2019} where the latent group structure in heterogeneous time-varying-coefficient panels is recovered via kernel estimation, hierarchical agglomerative clustering, and a generalized information criterion for the number of groups. In contrast to imposing group structure ex ante and estimating it directly as in \cite{Chen2019}, our framework treats grouping as a downstream step: test-implied dissimilarities feed into HAC clustering, yielding both a global decision on homogeneity and time-resolved evidence of heterogeneity.

% SIMUlATIONS AND EMPIRICAL APPLICATION

% rest of paper 
The remainder of the paper is organized as follows. Section \ref{sec:model} introduces the model. Section \ref{sec:test} presents our multiscale testing procedure, and Section \ref{sec:theory} establishes its theoretical properties. In Section \ref{sec:clust}, we discuss how our method can be extended to perform clustering and uncover unknown group structure in the data. %In Section \ref{sec:extensions}, we discuss how our method can be extended to test linear restrictions and perform clustering. Section \ref{sec:simulations} provides simulation results, while
In Section \ref{sec:application}, we apply our methods to an empirical dataset revisiting the relationship between U.S. monetary shock and $49$ foreign economies and U.S. GDP from \cite{Iacoviello2019}. Section \ref{sec:conclusion} concludes. Additional proofs and supplementary results are provided in Appendix.

% notation
Finally, some words on notation. %In order to simplify the calculations, we use the following notation: $\pr(\cdot| X = x):= \pr(\cdot| X_1 = x_1, \ldots, X_T = x_T)$, $\ex(\cdot|X = x) : = \ex(\cdot| X_1 = x_1, \ldots X_T = x_T)$ and $\var(\cdot|X=x) : = \var(\cdot| X_1=x_1, \ldots X_T=x_T)$.
For a vector $\mathbf{v} = (v_1,\ldots,v_D)\in \reals^D$, we write $||\mathbf{v}|| = \left(\sum^D_{i=1}v_i^2\right)^{1/2}$ and $||\mathbf{v}||_{\infty} = \max\{|v_1|, |v_2|, \ldots, |v_D|\}$ to denote its Euclidean and maximum ($\ell^\infty$) norm, respectively. %We use the shorthand notation $||\mathbf{v}|| = ||\mathbf{v}||_2$ for the $\ell_2$-norm.
Similarly, for a random vector $\mathbf{V}$, we denote its $\law^q$-norm for $q >0$ as $||\mathbf{V}||_{\law^q} = (\ex||\mathbf{V}||^q)^{1/q}$ (i.e., we always use the Euclidean base-norm). 
For a $D\times D$ square matrix $\bm A = (a_{d_1, d_2})_{1\leq d_1, d_2 \leq D}$ we denote induced norms $||\bm A||$ and $||\bm A||_{1}$ as $||\bm A|| = \sup \left\{\frac{||A\x||}{||\x||}: \x\in \reals^D, \x\neq \zeros\right\}$ and $||\bm A||_{1} = \max_{d_2=1, \ldots, D}  \sum_{d_1 = 1}^D |a_{d_1, d_2}|$, respectively. If $\bm A$ is positive semi-definite with eigenvalues $\lambda_1 \geq \ldots \geq \lambda_D$, we write $\rho_{\star}(\bm A) = \lambda_D$. For a set $\mathcal{S}$, we denote its cardinality as  $|\mathcal{S}|$. For a scalar $x$, we denote the greatest integer that is less or equal to $x$ by $\lfloor x \rfloor$. Throughout the paper, $C$ denotes a universal constant that may take different values on different lines. %Furthermore, for any sequence $a_n$, let $O_{pu}(a_n)$ and $o_{pu}(a_n)$ denote a random sequence that depends on some deterministic $x\in \mathcal{X}$ and that is $O_p(a_n)$ and $o_p(a_n)$, respectively, uniformly in $x$.

Furthermore, we introduce a general class of potentially nonlinear time series that can be defined as follows. Let $\eta_t$ for $t\in \integers$ be i.i.d. random variables. Define $\F_t = (\ldots, \eta_{t-1},\eta_t)$ and let $g : \reals^{\infty} \rightarrow \reals^D$ be a measurable function such that the random vector $g(\F_t)$ is well-defined. Following \cite{Wu2005} and later \cite{Karmakar2020}, we define the physical dependence measure of the process $g$ as
\begin{align}
    \delta_{q}(g,t) = ||g(\F_t) - g(\F'_t)||_{\law^q},
\end{align} 
where $\F_t^\prime = (\ldots, \eta_{-1}, \eta_0^\prime, \eta_1, \ldots, \eta_{t-1}, \eta_t)$ with $\eta^\prime_0$ being an i.i.d. copy of $\eta_0$. The quantity $\delta_{q}(g,t)$ measures the $t$-th lag dependence of the process $g$ in terms of $q$-th moment. For a detailed explanation of the physical dependence measure and its properties, we refer the reader to \cite{Wu2005}.

% \subsection*{Literature}

% \cite{Atak2023}

% \cite{Chen2019}

% \cite{Chen2021}

% \cite{Su2019}

% \cite{ZhouWu2010}

% KMT approximation (strong approximation theory): \cite{Xu2022} (weak dependence conditions on the product of factors and the shock, check Assumption F4), \cite{Karmakar2020} (Section 3.2 especially) with an extension by \cite{Mies2023} (almost the same results as by \cite{Karmakar2020} but with an explicit long-run covariance matrix estimation), maybe \cite{Cattaneo2022} (Sections 4.1 and 4.3), \cite{Li2020} (Theorem 2).

% Section 3.1, Theorem B.2 and Lemma B.3 in \cite{Karmakar2022}. They use the Gaussian approximation in \cite{Wu2011}.

% \cite{Cui2022}, Theorem 1, see also Equation 5.

% Notes after discussion with Wendun: 
% \begin{itemize}
%     \item this work is highly relevant to the literature testing slope heterogeneity. So far this literature has only considered static slopes. Some important references: \cite{Pesaran2008}, \cite{Su2013} and references therein.
%     \item testing only a single coefficient is economically relevant. A more general approach would be to consider $K$ regressors/slope coefficients and test them simultaneously. Depending on the difficulty, we could consider a uniform testing setting. 
% \end{itemize}
\section{The model framework} \label{sec:model}
In this section, we specify the model we consider in our paper and describe the preliminary step that are needed before constructing the test statistics. 
\subsection{The model}
Suppose that we observe a panel of $N$ time series $\mathcal{T}_i = \{(Y_{it}, \X^\prime_{t}): t = 1, \ldots, T\}$, of length $T$ where the responses $Y_{it}$ are real-valued random variables and the predictors $\X_t = (X_{t, 1}, \ldots, X_{t, D})^\prime \in \mathcal{X} \subset \reals^D$ are random vectors of covariates common for all time series (which holds, for instance, in factor models).  For the moment, we suppose that the set $\mathcal{X}$ is compact and $T\to \infty$, while $N$ and $D$ remain fixed.% and without loss of generality, we can consider $\mathcal{X} = [-1, 1]^D$.

We assume that each of the time series $\mathcal{T}_i$ follows a heterogeneous time-varying coefficients model:
% \begin{align}\label{eq:model}
% Y_{it} = \alpha_i + \X_{t}^\prime\bfbeta_i(t/T) + \varepsilon_{it}, \quad t = 1,\ldots, T,
% \end{align}
% where $\varepsilon_{it}$ are idiosyncratic error terms which can potentially be serially correlated across $t$ but are independent across $i$, $\alpha_i$ are unknown fixed effects which can exhibit cross-sectional dependence across $i$ and may be correlated with the predictors $\X_t$, and $\bfbeta_i = (\beta_{i, 1}, \ldots, \beta_{i, D})^\prime: [0, 1] \to \reals^D$ are unknown vector-valued coefficient functions that may be different among the units. Technical assumptions on the dependency structure in the model \eqref{eq:model} are discussed later in Section \ref{sec:theory}.
\begin{align}\label{eq:model}
Y_{it} = \X_{t}^\prime\bfbeta_i(t/T) + \varepsilon_{it}, \quad t = 1,\ldots, T,
\end{align}
where $\varepsilon_{it}$ are idiosyncratic error terms which can potentially be serially correlated across $t$ but are independent across $i$ and $\bfbeta_i = (\beta_{i, 1}, \ldots, \beta_{i, D})^\prime: [0, 1] \to \reals^D$ are unknown vector-valued coefficient functions that may be different among the units. Technical assumptions on the dependency structure in model \eqref{eq:model} are discussed later in Section \ref{sec:theory}. 

The main goal of this paper is to develop a rigorous statistical method that allows us to compare the unknown functions $\bfbeta_i$ across units, while providing information on where and when these differences occur. In what follows, we suppose that the functions $\bfbeta_i$ take the rescaled time $t/T$ as an argument which is common in the literature for nonparametric estimation of time-varying parameters \citep[e.g.,][]{Cai2007, Li2011}. Such rescaling means that, as the time dimension increases, we collect more and more data locally which ensures consistency of the nonparametric estimator that is used for our testing procedure. 

\subsection{Preliminary steps} \label{sec:estim}

We assume that $\bfbeta_i$ are sufficiently smooth coefficient functions satisfying Lipschitz continuity conditions. Formally, it can be formulated as follows:
\begin{enumerate}[label=(A\arabic*), leftmargin=1.05cm]
    \item \label{as:trends} For each $i= 1, \ldots, N$, the trend coefficient functions $\bfbeta_i$ are Lipschitz continuous on $[0,1]$ coordinate-wise, i.e., for all $d=1, \ldots, D$, $|\beta_{i, d}(v) - \beta_{i, d}(w)| \le C |v-w|$ for all $v,w \in [0,1]$.
\end{enumerate} 
We estimate $\bfbeta_i$ using a local constant regression approach. Specifically, for each fixed time point $u\in[0, 1]$, we minimize the locally weighted least squares criterion:
\begin{align}\label{eq:model:min}
    \sum^T_{t=1}\left(Y_{it} - \X_t'\bfbeta_i(u)\right)^2K\left(\frac{t/T - u}{h}\right).
\end{align}

Here, $K$ denotes a kernel function and $h$ is a bandwidth that controls the smoothness of the estimator. We assume that $h\in [h_T^{\min}, h_T^{\max}]$ with formal requirements on $h^{\min}_T$ and $h^{\max}_T$ stated further. Minimizing \eqref{eq:model:min} with respect to $\bfbeta_i$ results in the well-known Nadaraya-Watson estimator that has the following form:
\begin{align}\label{eq:model:beta_est}
\widehat{\bfbeta}_{i}(u, h) = \left[\frac{1}{\sqrt{Th}}\sum_{t=1}^T \X_{t} \X_{t}^\prime K\left(\frac{t/T - u}{h}\right)\right]^{-1}\frac{1}{\sqrt{Th}}\sum_{t=1}^T \X_{t} Y_{it} K\left(\frac{t/T - u}{h}\right).
\end{align}
Alternatively, we can estimate the unknown slope coefficients by a local linear approximation \citep[e.g.,][]{Li2011, Chen2012} or a sieve approximation \citep[e.g.,][]{Zhang2012, Atak2025}. However, the local constant approach allows us to use strong approximation theory for the errors $\varepsilon_{it}$ which guarantees valid critical values for our testing procedure.

In the rest of the paper, we employ the following shorthanded notation: \linebreak $K_{t, u, h}: = K\left((t/T - u)/h\right)$ for the value of the kernel function $K$ at $(t/T - u)/h$ and \linebreak $\bm M_{XKX}(u, h): = \frac{1}{\sqrt{Th}}\sum_{t=1}^T \X_{t} \X_t^\prime K_{t, u, h}$ for the (localized) design matrix. Using this notation, we can rewrite the estimator in \eqref{eq:model:beta_est} as 
\begin{align*}
\widehat{\bfbeta}_{i}(u, h) = \bm M_{XKX}^{-1}(u, h)\left(\frac{1}{\sqrt{Th}}\sum_{t=1}^T \X_{t} Y_{it} K_{t, u, h}\right).
\end{align*}

% \begin{remark}
%     Our model allows for a constant term to be present among the covariates $\X_t$. In this case, the model can be re-written as having an unknown deterministic unit-specific trend function which is more general than having fixed effects.
% \end{remark}

\begin{remark}  
    Introducing time-varying coefficient functions in \eqref{eq:model} generalizes the model in \cite{KhismatullinaVogt2022}, which has constant coefficients $\bfbeta_i$ and a deterministic trend function. The authors provide a multiscale test for (local) differences in the trend functions. We allow all coefficients to be time-varying, where $X_{t,1}$ does not necessarily need to be constant.
\end{remark}
\section{Testing procedure}\label{sec:test}
This section presents the multiscale testing procedure for detecting heterogeneity in regression coefficients. The primary objective is to assess whether the time-varying coefficient functions $\bfbeta_i$ in model \eqref{eq:model} are the same across all time series. Unlike conventional tests that focus on global comparison, our approach captures localized variations by testing over multiple time scales simultaneously.

\subsection{Hypothesis formulation}\label{sec:hypothesis}
We formally define the null hypothesis as:
\begin{align}\label{eq:hypothesis}
    H_0 : \bfbeta_1(u) = \bfbeta_2(u) = \ldots = \bfbeta_N(u) \quad \forall u\in[0, 1],
\end{align}
which states that the coefficient functions $\bfbeta_i$ are everywhere identical across all units $i=1, \ldots, N$. Under $H_0$, model \eqref{eq:model} simplifies to a homogeneous linear panel data model. In this setting, pooling data across units yields an efficient estimator of the common slope function. However, when $H_0$ is violated, pooling can lead to biased estimates and incorrect inference due to underlying coefficient heterogeneity \citep[see, e.g.,][]{Hsiao1997}. Testing $H_0$ is therefore critical for ensuring valid estimation and inference in time-varying panel models.

Several approaches exist for testing $H_0$. Traditional parametric methods, such as the Wald test and the likelihood ratio test, compare restricted and unrestricted estimators. More recently, \cite{Atak2025} proposed a nonparametric test based on residuals obtained from the estimation procedure under $H_0$. However, while these tests are helpful in detecting slope heterogeneity in general, they mostly focus on global departures from homogeneity and do not identify where and when coefficient differences occur.

In this section, we describe a multiscale testing framework that provides richer insights than standard methods. Following the perspective of \cite{KhismatullinaVogt2020, KhismatullinaVogt2022, KhismatullinaVogt2023}, our approach is designed to answer three key questions:
\begin{enumerate}
    \item Is the null hypothesis $H_0$ violated? That is, does at least one coefficient function differ somewhere across units?
    \item If the null hypothesis $H_0$ is not true, then which coefficient functions differ?
    \item At which time intervals do these differences arise? %Detecting local deviations rather than relying on a single global comparison.
\end{enumerate}

To achieve this, we introduce a series of local null hypotheses, where "local" refers to a certain pair of units and a specific time intervals rather than the entire time domain. We then construct test statistics that allow for separate evaluation of each local null hypothesis while accounting for multiple testing problem at the same time. The core idea is to aggregate the conclusions from these local tests into a unified confidence statement, providing a statistically rigorous framework for detecting heterogeneity in time-varying coefficients.

We proceed as follows. For any interval $[u-h,u+h] \subseteq [0,1]$ (where $u$ denotes the location and $h$ the bandwidth) and for any pair of units $1 \leq i < j \leq N$, consider the local null hypothesis
\[ H_0^{[i,j]}(u,h): \bfbeta_i(w) = \bfbeta_j(w) \text{ for all } w \in [u-h,u+h], \] that states that the coefficient functions of units $i$ and $j$ are identical on the interval $[u-h,u+h]$. Reformulating the global null hypothesis in terms of these local hypotheses gives:
\begin{align*}
H_0: \ & \text{The hypotheses } H_0^{[i,j]}(u,h) \text{ hold true for all intervals } [u-h,u+h] \subseteq [0,1] \\ & \text{ and for all } 1 \le i < j \le N. 
\end{align*} 

Thus, instead of testing a single global null hypothesis $H_0$, our method assesses a collection of local ones, enabling us to identify specific deviations in coefficient functions both across units and over time. Moreover, the multiscale perspective ensures that our test is sensitive to both long-term trends and short-term structural changes.

\subsection{Construction of the test statistic}

The main idea of our multiscale method is to test simultaneously hypotheses $H_0^{[i,j]}(u,h)$ for all intervals $[u-h, u+h]\subseteq[0, 1]$ and all pairs $(i,j)$ with $1\leq i < j \leq N$. In practice, we are unable to consider all possible intervals $[u-h,u+h] \subseteq [0,1]$, and we need to work with a countable subset of these intervals. We consider the following set. We define a collection of location-bandwidth points $\mathcal{G}_T = \{(u, h): [u-h, u+h] \subseteq[0, 1] \text{ with } u = t/T, h = s/T \text{ for some } t, s = 1, \ldots, T \text{ and } h\in [h^{\min}_T, h^{\max}_T]\}$. Here, as before, $h^{\min}_T$ and $h^{\max}_T$ denote the minimal and maximal bandwidths, respectively, with specific requirements on their selection provided below in Assumption \ref{as:GT_bound}. The set $\grid$ is chosen to be sufficiently rich so that the resulting intervals collectively cover the entire unit interval. We allow $\mathcal{G}_T$ to be quite large compared to the length of the time series $T$. Formal assumptions on $\mathcal{G}_T$, $h^{\min}_T$ and $h^{\max}_T$ are given in Section \ref{sec:as}.

Now we are ready to construct the localized pairwise test statistics $\hat{S}_{ij}(u,h)$. For a given pair $(u, h) \in \mathcal{G}_T$ and all $i=1, \ldots, N$, we estimate $\bfbeta_i(u)$ using the nonparametric kernel estimator $\widehat{\bfbeta}_{i}(u, h)$ as defined in \eqref{eq:model:beta_est}. Based on these estimates, for each pair $(i, j)$, $1 \leq i<j \leq N$ we consider the following quantity: 
\begin{align*}
    \hat{S}_{ij}(u,h) = \Big|\Big| \widehat{\bm \Sigma}_{ij}^{-1/2} \bm M_{XKX}(u, h)  \left(\hat{\bfbeta}_{i}(u, h) - \hat{\bfbeta}_{j}(u, h)\right)\Big|\Big|_{\infty},
\end{align*}
where $\widehat{\bm \Sigma}_{ij} = (\widehat{\bm \Sigma}_{i} + \widehat{\bm \Sigma}_{j})/2$ with $\widehat{\bm \Sigma}_{i}$ and $\widehat{\bm \Sigma}_{j}$ represent appropriate estimators of the long-run covariance matrices ${\bm \Sigma}_{i}$ and ${\bm \Sigma}_{j}$ of the processes $\{\X_t \varepsilon_{it}\}_{t=1}^T$ and $\{\X_t \varepsilon_{jt}\}_{t=1}^T$, respectively. Our procedure requires an estimator $\widehat{\bm \Sigma}_{i}$ to have the rate of convergence $\rho_T = o(\log T)$ which is easily achieved using most standard estimators, for example, HAC estimator as proposed in \cite{Andrews1991}. Discussion of the estimation procedure for ${\bm \Sigma}_{i}$ that we follows in the simulation study and in the empirical analysis is provided in Section \ref{sec:lrv_estimation}.

Heuristically, the statistic $\hat{S}_{ij}(u,h)$ approximates the distance between $\bfbeta_i$ and $\bfbeta_j$ over the interval $[u-h, u+h]$. To ensure scale invariance, we normalize the test statistic using the (localized) variance of $\X_{t}$ through $\bm M_{XKX}(u, h)$ and the estimated variances of the processes $\{\X_{t} \varepsilon_{it}\}_{t=1}^T$ and $\{\X_{t} \varepsilon_{jt}\}_{t=1}^T$. By construction, $\hat{S}_{ij}(u, h)$ is always nonnegative, with smaller values indicating that the difference between $\bfbeta_i$ and $\bfbeta_j$ over the given interval is quite small. In contrast, large positive values of $\hat{S}_{ij}(u,h)$ suggest substantial differences, making the rejection of the local null hypothesis $H_0^{[i, j]}(u, h)$ more likely.

To test the global null hypothesis $H_0$, we aggregate the local statistics $\hat{S}_{ij}(u,h)$ over all location-bandwidth points $(u,h)\in \mathcal{G}_T$ and all pairs $(i,j), 1 \leq i<j \leq N$, using the aggregation scheme proposed by \cite{DuembgenSpokoiny2001}. The resulting multiscale test statistic is given by
\begin{align}\label{eq:test_stat}
\widehat{\Psi}_{T} = \max_{1 \le i < j \le N} \max_{(u,h)\in \mathcal{G}_T} \big\{ \hat{S}_{ij}(u,h) - \lambda(h) \big\},
\end{align}
where $\lambda(h)$ are (appropriately chosen) additive correction terms. %Unlike a simple supremum over $\hat{S}_{ij}(u, h)$, our aggregation approach applies a correction function $\lambda(h)$ to improve finite-sample performance.
In brief, $\lambda(h)$ represents a term designed to account for the multiple testing problem inherent in our framework. The functional form of $\lambda(h)$ does not affect the asymptotic convergence results for our test statistic but may affect the finite sample performance. In our work, $\lambda(h)$ is determined by the extreme value theory and depends on the bandwidth $h$: drawing on the results from \cite{DuembgenSpokoiny2001}, the correction term for our setting is given by $\lambda(h) = \sqrt{2\log(1/2h)}$.

\subsection{Computation of the critical values} \label{sec:computation}

Suppose for a moment that the exact $(1-\alpha)$-quantile $q_{T}^*(\alpha)$ of the multiscale test statistic $\widehat{\Psi}_{T}$ under $H_0$ were known. The multiscale test would then follow the rejection rule: 
\begin{itemize}[leftmargin=1cm]
\item[(T$^*$)] Reject $H_0$ if $\widehat{\Psi}_{T} > q_{T}^*(\alpha)$. 
\end{itemize}
By construction, this decision rule ensures a rigorous level-$\alpha$-test, meaning that under $H_0$ we have $\mathbb{P}(\widehat{\Psi}_{T} > q_{T}^*(\alpha)) = \alpha$. 

%To control the overall significance level of our multiscale test while accounting for the simultaneous testing of many local null hypotheses, we need to account for the multiple testing problem when constructing the critical values. In theory, the global null hypothesis $H_0$ should be rejected if the aggregated test statistic $\widehat{\Psi}_T$ exceeds a suitable quantile threshold. However,
In practice, however, $q_{T}^*(\alpha)$ is unknown and can not be computed analytically due to the complicated dependence structure of the individual test statistics across various location-bandwidth points. Hence, it cannot be used to set up the test. In what follows, we show how to construct an asymptotic approximation $q_{T}(\alpha)$ that is both computationally feasible and statistically valid. In particular, we require that under $H_0$, $q_{T}(\alpha)$ satisfies the asymptotic property 
\begin{equation}\label{q-approx}
\mathbb{P} (\widehat{\Psi}_{T} > q_{T}(\alpha)) = \alpha + o(1).
\end{equation}

We use strong approximation theory to compute the critical values $q_{T}(\alpha)$, and in Section~\ref{sec:theory}, we show that this approach ensures an asymptotically valid testing procedure. 

For calculation of the critical values, we do the following. Under the null $H_0$, the individual test statistics
%suppose that we could perfectly estimate $\bm \Sigma_i$ so that $\widehat{\bm \Sigma}_i = \bm \Sigma_i$ for all $i$. Using the notation $\bm \Sigma_{ij} = (\bm \Sigma_i + \bm \Sigma_j)/2$, the individual test statistics $\hat{S}_{ij}(u, h)$ is then given by
\begin{align*}
    \hat{S}_{ij}(u, h)&=\Big|\Big|\widehat{{\bm \Sigma}}_{ij}^{-1/2}\bm M_{XKX}(u, h)\left(\hat{\bfbeta}_{i}(u, h) - \hat{\bfbeta}_{j}(u, h)\right)\Big|\Big|_{\infty} \\   
    %&=\left(T^{-1}\sum^T_{t=1}\X_tY^0_{it}K(\frac{t/T-u}{h}) - T^{-1}\sum^T_{t=1}\X_tY^0_{jt}K(\frac{t/T-u}{h})\right)'\Sigma^{-1}_{ij}\left(T^{-1}\sum^T_{t=1}\X_tY^0_{it}K(\frac{t/T-u}{h}) - T^{-1}\sum^T_{t=1}\X_tY^0_{it}K(\frac{t/T-u}{h})\right)\\
    &= \Bigg|\Bigg|\widehat{\bm\Sigma}^{-1/2}_{ij}\left(\frac{1}{\sqrt{Th}}\sum^T_{t=1}\X_t\big(Y_{it} - Y_{jt}\big)K_{t, u, h}\right) \Bigg|\Bigg|_{\infty}\\
    &= \Bigg|\Bigg|\widehat{\bm\Sigma}^{-1/2}_{ij}\Bigg(\frac{1}{\sqrt{Th}}\sum^T_{t=1}\X_t\bigg(\X_{t}^\prime\bfbeta_i(t/T) + \varepsilon_{it} - \X_{t}^\prime\bfbeta_j(t/T) - \varepsilon_{jt} \bigg)K_{t, u, h}\Bigg) \Bigg|\Bigg|_{\infty},
\end{align*}
is exactly equal to
\begin{align}\label{eq:individ_under_null}
    &\hat{S}^0_{ij}(u,h) = \Bigg|\Bigg|\widehat{\bm\Sigma}^{-1/2}_{ij}\left(\frac{1}{\sqrt{Th}}\sum_{t=1}^T \X_{t} \left(\varepsilon_{it}  - \varepsilon_{jt}\right) K_{t, u, h}\right)\Bigg|\Bigg|_{\infty}.
\end{align}
Under some regularity conditions and a ceratin rate of consistency of the estimators $\widehat{{\bm \Sigma}}_{i}$ and $\widehat{{\bm \Sigma}}_{j}$ of ${\bm \Sigma}_{i}$ and ${\bm \Sigma}_{j}$, respectively, we can show that $\hat{S}^0_{ij}(u,h)$ can be approximated by
\begin{align*}
    S_{ij}(u,h)= \Bigg|\Bigg|\bm \Sigma^{-1/2}_{ij}\left(\frac{1}{\sqrt{Th}}\sum^T_{t=1}\big(\bm\Sigma^{1/2}_i \Z_{it}  - \bm\Sigma_j^{1/2}\Z_{jt}\big)K_{t, u, h}\right)\Bigg|\Bigg|_{\infty},
\end{align*}

where $\Z_{it}$ are i.i.d. across $i$ and $t$ multivariate standard normal random vectors that are independent of $\X_s$ and $\varepsilon_{js}$. Essentially, when constructing $S_{ij}(u, h)$, we have replaced the terms $\X_t\varepsilon_{it}$ by the scaled multivariate Gaussian random vectors $\bm \Sigma^{1/2}_i \Z_{it}$. Under an additional simplifying assumption of $\bm \Sigma_i = \bm\Sigma_j$, the Gaussian version $S_{ij}(u, h)$ of the test statistic can be further reduced to
\begin{align} \label{eq:gauss_equal_var}
	S_{ij}(u,h)&=\Big|\Big|\frac{1}{\sqrt{Th}}\sum^T_{t=1}\big(\Z_{it} - \Z_{jt}\big)K_{t, u, h}\Big|\Big|_{\infty}.
\end{align}
Note that in this case, $S_{ij}(u, h)$ does not depend on any unknown quantities and its quantiles can be (approximately) calculated via Monte Carlo simulations.

In order to construct a global test statistic, we aggregate $S_{ij}(u, h)$ in a similar way as $\hat{S}_{ij}(u, h)$:
\begin{align} \label{eq:gauss_stat}
    \Phi_{T} = \max_{1\leq i<j\leq N}\max_{(u,h)\in \mathcal{G}_T}\{S_{ij}(u,h) - \lambda(h)\}.
\end{align}

As we will rigorously prove in Section \ref{sec:theory}, the $1-\alpha$ quantile $q^*(\alpha)$ of the aggregated data-dependent test statistic $\widehat{\Psi}_{T}$ can be approximated by the $1-\alpha$ quantile $q^*(\alpha)$ of the similarly aggregated Gaussian test statistic $\Phi_{T}$. Equation \eqref{eq:gauss_equal_var} shows that the distribution of $S_{ij}(u,h)$ for each pair $(i, j), 1\leq i< j \leq N$, and each $(u, h)\in\grid$, and thus of $\Phi_T$ depends only on the Gaussian random vectors $\Z_{it}$. Therefore, accurate critical values of $\Phi_T$ can be computed by simulating draws of $\Z_{it}$ for a large number of times $B$. In our simulation study and application analysis, we choose $B = 5000$, but we have repeated the analysis with $B = 1000$ and the results are roughly the same.

Here we formally describe our testing approach. For a given significance level $\alpha\in(0,1)$, we can compute the critical values of $\Phi_T$ according to the following steps:
\begin{enumerate}[label = {Step \arabic*.},align=left]
    \item Simulate $B$ independent draws of a matrix
    $$\mathcal{Z}^{b} = (\Z^b_{it})_{i=1, \ldots, N; t=1, \ldots, T},  
%    \begin{pmatrix}
%        \Z^{b}_{11} &\ldots &\Z^{b}_{1T}\\
%    \Z^{b}_{21} &\ldots &\Z^{b}_{2T} \\
%    \vdots &\ddots &\vdots \\
%    \Z^{b}_{N1} &\ldots &\Z^{b}_{NT}
%    \end{pmatrix},
\quad b = 1, 2, \ldots, B, 
    $$
where each entry $\Z^b_{it}$ is an independent multivariate standard normal random variable of dimension $D$.
\item Compute the simulated individual test statistic
%    {\scriptsize\begin{align*}
%        S^b_{ij}(u,h)&= \left(\frac{1}{T}\sum^T_{t=1}\big(\bm\Sigma^{\frac{1}{2}}_i(\Z^b_{it} - \bar{\Z}^b_i) - \bm\Sigma_j^{\frac{1}{2}}(\Z^b_{jt} - \bar{\Z}^b_j)\big)K_{t, u, h}\right)^\prime\bm\Sigma^{-1}_{ij}\left(\frac{1}{T}\sum^T_{t=1}\big(\bm\Sigma^{\frac{1}{2}}_i(\Z^b_{it} - \bar{\Z}^b_i) - \bm\Sigma_j^{\frac{1}{2}}(\Z^b_{jt} - \bar{\Z}^b_j)\big)K_{t, u, h}\right),
%    \end{align*}}
%    
%    which under the assumption that $\bm\Sigma_i = \bm\Sigma_j$ does not depend on the unknown quantities and can be calculated as  
\begin{align*}
    S^b_{ij}(u,h)&= \Big|\Big|\frac{1}{\sqrt{Th}}\sum^T_{t=1}\big(\Z^b_{it}  - \Z^b_{jt}\big)K_{t, u, h}\Big|\Big|_{\infty}.
\end{align*}
    
    \item Aggregate the simulated statistics over all pairs $(i, j), 1 \leq i < j \leq N$, and all location-bandwidth points $(u, h)\in \mathcal{G}_T$ as
    \[ \Phi^{b}_{T} = \max_{1 \le i < j \le N} \max_{(u,h)\in \mathcal{G}_T} \big\{ S_{ij}^b(u,h) - \lambda(h) \big\}.\]
    \item Calculate the critical value $q_T(\alpha)$ as the empirical $(1-\alpha)$-quantile of the simulated values $\left\{\Phi^{b}_{T}\right\}^B_{b=1}$.
\end{enumerate}

This resampling approach ensures that the estimated critical values correctly approximate the unknown quantile $q^*_T(\alpha)$, allowing for valid inference. The computational complexity of this procedure is manageable even for large samples, as the Monte Carlo step can be computed in parallel and does not require re-estimation of the coefficient functions.

\subsection{The testing procedure} \label{sec:testprocedure}
For a given significance level $\alpha\in(0,1)$, we conduct the multiscale test to assess the global null hypothesis $H_0$ using the following decision rule:
\begin{itemize}[leftmargin=0.8cm]
\item[(T)] Reject $H_0$ if $\widehat{\Psi}_{T} > q_{T}(\alpha)$. 
\end{itemize}

By adopting this criterion, the proposed method serves as a test for the overall hypothesis $H_0$. Alternatively, it can be interpreted as a simultaneous test for the family of local null hypotheses $H_0^{[i,j]}(u,h)$ across all location-bandwidth points $(u,h)\in \mathcal{G}_T$ and all unit pairs $(i,j)$, $1 \leq i < j \leq N$. This interpretation allows us to formulate a local decision rule:
\begin{itemize}[leftmargin=1.5cm]
\item[(T$_\text{mult}$)] 
For each interval $[u-h,u+h]$ with $(u,h)\in \mathcal{G}_T$ and for each pair $(i, j)$ with \linebreak $1 \leq i<j\leq N$, reject $H_0^{[i,j]}(u,h)$ if $\hat{S}_{ij}(u,h) - \lambda(h) > q_{T}(\alpha)$. 
\end{itemize}
This local rule ensures that hypothesis testing is performed across multiple time intervals while our approach for constructing of the critical values simultaneously controls for resulting multiple comparisons.

A crucial property of the multiscale test is that it enables rigorous confidence statements about where and when significant deviations occur. Specifically, in Section \ref{sec:theory}, we prove the following result under suitable regularity conditions:
\begin{itemize}[leftmargin=0.8cm]
\item[] \textit{With asymptotic probability at least $1-\alpha$, the hypothesis $H_0^{[i,j]}(u,h)$ is violated for all pairs $(i,j)$ and for all intervals $[u-h,u+h]$ satisfying $\hat{S}_{ij}(u,h) - \lambda(h) > q_{T}(\alpha)$.} 
\end{itemize}

This result allows us to make the following simultaneous confidence statement:
\begin{itemize}[leftmargin=0.8cm]
\item[] \textit{With at least $(1- \alpha)$ (asymptotic) confidence, we can claim that the hypothesis $H_0^{[i,j]}(u,h)$ is violated for all pairs of time series $(i,j)$ and for all intervals \linebreak $[u-h,u+h]$ for which our test rejects.}
\end{itemize}
This provides a statistically robust way to detect heterogeneous time-varying coefficient functions across unit pairs. By accumulating local test results, we construct a comprehensive confidence region for coefficient differences over time and for all pairwise comparisons, adding interpretability to the statistical inference.

\subsection{Long-run covariance estimation} \label{sec:lrv_estimation}
Define $\vv_{it}=\X_t\varepsilon_{it}$. As was mentioned before, our multiscale test relies on a consistent estimator of the long-run variance $\bm\Sigma_i = \sum_{\ell=-\infty}^{\infty}\cov(\vv_{i0}, \vv_{i\ell})$. In this section, we describe one of the possible estimators.

We can rewrite the long-run variance $\bm\Sigma_i$ as follows:
\begin{align*}
    \bm \Sigma_i &= \lim_{T\rightarrow\infty}\sum^{T-1}_{\ell=-T+1}\bm \Gamma_{i,T}(\ell), \text{ where}&\\
    \bm \Gamma_{i,T}(\ell) &= \begin{cases}
        \frac{1}{T}\sum^T_{t=\ell+1}\ex(\vv_{it}\vv_{i(t-\ell)}'), &\ell\geq 0,\\
        \frac{1}{T}\sum^T_{t=-\ell+1}\ex(\vv_{i(t+\ell)}\vv_{it}'), &\ell<0.
    \end{cases}
\end{align*}
A popular choice in the time series literature is the kernel heteroskedasticy and autocorrelation consistent (HAC) covariance estimator. The idea of the kernel HAC estimator is to estimate autocovariances $\bm \Gamma_{i,T}(\ell)$ by sample autocovariances $\widehat{\bm\Gamma}_{i,T}(\ell)$ and to attach lower weights to autocovariances with large $|\ell|$ using a kernel function. The kernel weights smooth the sample autocovariances, ensuring a positive semi-definite covariance estimator.

Specifically, we proceed as follows. Let $\kappa$ denote a real-valued covariance kernel function with some fixed bandwidth $\chi$. The class of fixed-bandwidth kernel HAC estimators is given by
\begin{align}
    \widehat{\bm \Sigma}_i(\chi) &= \frac{T}{T-D}\sum^{T-1}_{\ell=-T+1} \kappa\left(\frac{\ell}{\chi}\right)\widehat{\bm \Gamma}_{i,T}(\ell)\\
    \widehat{\bm \Gamma}_{i,T}(\ell) &= \begin{cases}
        \frac{1}{T}\sum^T_{t=\ell+1}\hat{\vv}_{it}\hat{\vv}_{i(t-\ell)}', \quad \ell\geq 0,\\
        \frac{1}{T}\sum^T_{t=-\ell+1}\hat{\vv}_{i(t+\ell)}\hat{\vv}_{it}', \quad \ell<0,
    \end{cases}
\end{align}
where $\hat{\vv}_{it} = \X_t \hat{\varepsilon}_{it}$ with $\hat{\varepsilon}_{it}$ being the residuals after plugging in some estimator $\hat{\bfbeta}_i(t/T)$ of $\bfbeta_i(t/T)$. Importantly, the covariance kernel function $\kappa$ should give a smaller weight to $\widehat{\bm \Gamma}_{i,T}(\ell)$ as $|\ell|$ increases. The following assumptions guarantee that the kernel HAC covariance estimator $\widehat{\bm \Sigma}_i$ converges to a true long-run variance matrix $\bm \Sigma_i$. The assumptions are a simplified version of Assumptions A-C in \cite{Andrews1991} and Assumptions 5.1-5.3 in \cite{Belotti2023} where we used the fact that we consider a special case of a linear model compared to a general one discussed in these papers.

Let $m^{(abcd)}_i(t,t+r,t+s,t+o)$ denote the time-$t$ fourth-order cumulant of \linebreak $\left(\vv^{(a)}_{it}, \vv^{(b)}_{i(t+r)}, \vv^{(c)}_{i(t+s)}, \vv^{(d)}_{i(t+o)}\right)$ where $\vv^{(a)}_{it}$ is the $a$th element of a vector $\vv_{it}$. That is,
\begin{align*}
    &m_i^{(abcd)}(t,t+r,t+s,t+o)=\ex\left(\vv^{(a)}_{it}\vv^{(b)}_{i(t+r)}\vv^{(c)}_{i(t+s)}\vv^{(d)}_{i(t+o)}\right) - \ex\left(\tilde{\vv}^{(a)}_{it}\tilde{\vv}^{(b)}_{i(t+r)}\tilde{\vv}^{(c)}_{i(t+s)}\tilde{\vv}^{(d)}_{i(t+o)}\right),
\end{align*}
where $\{\tilde{\vv}_{it}\}_{t=1}^T$ is a Gaussian sequence with the same mean and covariance structure as $\{\vv_{it}\}_{t=1}^T$. Now, we can formulate the requirements on the dependence structure of the process $\{\vv_{it}\}_{t=1}^T$ as follows. %Furthermore, for each unit $i$, we denote the smoothness of the corresponding spectral density matrix $f_i(\lambda)$ at $\lambda=0$ by
%\begin{align*}
%    f_i^{(q)} = \frac{1}{2\pi} \sum^{\infty}_{\ell=-\infty}|j|^q\bm \Gamma_{i,T}(\ell) \qquad q\in[0,\infty).
%\end{align*}
%We denote $\kappa_q$ as a measure of smoothness, following \cite{}, which is defined as
% \begin{align*}
%     \kappa_q = \lim_{x\rightarrow0}\frac{1-\kappa(x)}{|x|^q} \qquad q\in[0,\infty).
% \end{align*}

\begin{enumerate}[label=(A\arabic*), leftmargin=1.05cm]
\setcounter{enumi}{1} 
\item \label{as:hac_newey} For all $i=1,\ldots,N$, $T^{\vartheta}(\hat{\bfbeta}_i - \bfbeta_i) = O_p(1)$ for some $\vartheta\in (0,1/2)$, \linebreak $\sup_{t\geq 1}\ex(||\vv_{it}||^2) < \infty$, and
$\sup_{t\geq 1}\ex(||\X_{t}\X_{t}'||^2)  <\infty$. Moreover, $\int |\kappa(y)|dy < \infty$.
\item \label{as:hac_stationary} For all $i=1,\ldots,N$, the process $\{\vv_{it}\}_{t=1}^T$ is a mean-zero, fourth-order stationary sequence with $\sum^{\infty}_{\ell=-\infty} ||\bm \Gamma_{i, T}(\ell)|| < \infty$ and $$\sum^{\infty}_{k=-\infty}\sum^{\infty}_{s=-\infty}\sum^{\infty}_{o=-\infty}|m_i^{(abcd)}(0,k,s,o)| < \infty$$ for all $a,b,c,d = 1, \ldots, D$.
\item \label{as:hac_stationary_joint} For all $i=1,\ldots,N$, Assumption \ref{as:hac_stationary} holds with $\vv_{it}$ replaced by \linebreak $(\vv_{it}', (\X_t\X_t' - \ex\left(\X_t\X_t'\right))')'$.
% \item \label{as:hac_bartlett} $k(u) = 1-|u|$ for $|u|\leq 1$ and $0$ otherwise, and the rate $\ell = \ell(T) = O(1/\rho_T)$ with $\rho_T = T^{-1/3}$.
\end{enumerate}
\cite{Belotti2023} establish that under Assumptions \ref{as:hac_newey} - \ref{as:hac_stationary_joint}, using nonparametric estimator $\hat{\bfbeta}_i$ of $\bfbeta_i$ with data-driven bandwidths leads to the rate of convergence of a kernel HAC estimator to be polynomial in $T$. In the appendix, we provide a similar result for the rate of convergence for a fixed bandwidth parameter. This holds true for various kernel functions, and since the necessary condition for our multiscale procedure is to have a rate of convergence $\rho_T = o(\log T)$, our main results remain valid for a number of HAC specifications. 
 %For example, the Bartlett kernel \cite{Newey1987} achieves convergence with rate $\rho_T = T^{-1/3}$ to $\bm \Sigma$ for all $i$, see also \cite{Andrews1991}. Note that the rate $\rho_T$ is different for other kernel choices. Our results remain unchanged for different choices, since a polynomial rate is sufficient.
%\begin{lemma}
%    Let Assumptions \ref{as:hac_newey} - \ref{as:hac_stationary_joint} hold. Moreover, it holds that $\chi^{2q+1}/T \rightarrow \gamma \in (0,\infty)$ for some $q\in(0,\infty)$, and that $||f^{(q)}||<\infty$. Then, $\hat{\Sigma}_i(\chi)\rightarrow \Sigma_i$ for all $i$. \commentsB{I didn't write down the proof, but seems straightforward following \cite{Andrews1991} and \cite{Belotti2023}}
%\end{lemma}

In the estimation procedure described above, we need to choose the bandwidth value which determines the truncation of the sample autocovariances. A common practice which we follow in the simulation study and in the empirical analysis is to set $\chi \approx T^{1/3}$. Alternatively, one can opt for the selection procedure proposed by \cite{Newey1994}, or the parametric selection methods by \cite{Andrews1991} and \cite{Andrews1992}.

% \begin{remark}
% Under weak dependence, we can use the Bartlett estimator as a proxy for $\Sigma_i$ with some window length $\omega(T) = log_{10}(T)$. This estimator $\hat{\Sigma}_i$ will satisfy $|\hat{\Sigma}_i - \Sigma_i|_E = o_p(1)$ as $T\rightarrow \infty$, see \cite{aue2009break} and \cite{andrews1991heteroskedasticity}.
% \end{remark}
\section{Theory}\label{sec:theory}
In this section, we present the theoretical results that are necessary for validity of our multiscale testing procedure.

\subsection{Assumptions}\label{sec:as}
To establish the theoretical properties of our test, we impose a set of assumptions about the behavior of the regressors, errors, and underlying dependence structure in the model.

\begin{enumerate}[label=(A\arabic*), leftmargin=1.05cm]
\setcounter{enumi}{4}
    \item\label{as:err} Processes $\bfvareps_i = \{\varepsilon_{it}\}_{t=1}^T$ are independent across $i$.
%    \item\label{as:x-err1} The processes $\{\varepsilon_{it}\}^T_{t=1}$ are martingale difference sequences with respect to the filtration $\F_{it}\sigma(\eta_{it},\ldots,\eta_{it})$, where the variables $\eta_{is}$ are i.i.d. across $s$. Furthermore, the variables $\varepsilon_{it}$ satisfy $\ex[\varepsilon_{it}| \X_1, \ldots, \X_T] = 0$,  $\ex[\varepsilon^2_{it}|\X_1, \ldots, \X_T] = \sigma^2_t$ with $T^{-1}\sum^T_{t=1}\sigma^2_t = O(1)$, and $\ex[\varepsilon_{it}\varepsilon_{is}|\X_1, \ldots, \X_T] = \sigma_{t, s}$ with $\sum^T_{s=1}|\sigma_{t,s}| = O(1)$. The variables $\vv_{it} = \X_t \varepsilon_{it} \in \reals^D$  allow for the representation $\vv_{it} = g_i(\F_{it})$ where $\F_{it} = (\ldots,\eta_{i(t-1)}, \eta_{it})$ and $\eta_{is}$ are i.i.d. across $s$.
    \item\label{as:x-err1} For each $i = 1, \ldots, N$, the process $\bfvareps_i$  satisfies $\ex[\varepsilon_{it}| \X_1, \ldots, \X_T] = 0$, \linebreak $\ex[\varepsilon^2_{it}|\X_1, \ldots, \X_T] = \sigma^2_{\varepsilon}$ and $\ex[\varepsilon_{it}\varepsilon_{is}|\X_1, \ldots, \X_T] = \sigma_{\varepsilon, \varepsilon}$ for all $t, s = 1, \ldots, T$. 

    \item\label{as:x-err2} For each $i = 1, \ldots, N$, the variables $\vv_{it} = \X_t \varepsilon_{it} \in \reals^D$  allow for the representation $\vv_{it} = g_i(\F_{it})$ where $\F_{it} = (\ldots,\eta_{i(t-1)}, \eta_{it})$ and $\eta_{is}$ are i.i.d. across $s$.

    \item\label{as:x-err3} For each $i = 1, \ldots, N$, it holds that $\ex(||\vv_{it}||^q) \leq C < \infty$ for some $q>4$
    
    \item\label{as:x-err4} For each $i$, $\sum_{s\geq t} \delta_q(g_i, s) = O(t^{-\xi}(\log t)^{-A})$, where
    \begin{align*}
        A &> \frac{(2q+q^2)\xi + q^2 + 3q + 2 + \gamma^{1/2}}{q(1+q+2\gamma)},\\
        \gamma &= q^2(q^2 + 4q -12)\xi^2 + 2q(q^3+q^2-4q-4)\xi + (q^2-q-2)^2,\\
        \xi &>\xi_0 \frac{q^2 - 4 + (q-2)\sqrt{q^2 + 20q + 4}}{8q},
    \end{align*}
    and $q$ is defined in \ref{as:x-err3}.
    \item\label{as:x-err5} There exists $\lambda_*>0$ and $l_*\in \naturals$ such that for all $i$, $t\geq 1$ and $l\geq l_*$, the smallest eigenvalue of $\var(\sum^{t+l}_{s=t+1}\vv_{is})$ is sufficiently bounded away from zero, i.e.,
    \begin{align*}
        \rho_{\star}\left(\var\left(\sum^{t+l}_{s=t+1}\vv_{is}\right)\right)\geq \lambda_* l.
    \end{align*}

%    \item\label{as:eig_xx}The smallest eigenvalue of $\left(\sum^T_{t=1}\X_t \X_t' K(\frac{t/T-u}{h})\right)$ is bounded away from zero. \textcolor{red}{With high probability, right?}
    \item\label{as:eig_xx} For all $(u,h)$, it holds that $\bm M_{XKX}(u, h) = \frac{1}{\sqrt{Th}}\sum^T_{t=1}\X_t \X_t^\prime K_{t, u, h}$ is positive definite with probability going to 1.
\item \label{as:GT_bound}$|\mathcal{G}_T| = O(T^{\theta})$ for some arbitrary large but fixed constant $\theta>0$. Furthermore, we assume that $h^{\min}_T \gg T^{2/(3q) - 1/3}$ where $q$ is defined in \ref{as:x-err2} and $h^{\max}_T \leq 1/4$.
\item \label{as:kernel} The kernel $K$ is non-negative, symmetric about zero, integrates to one and there exists $\delta_K>0$ and $c_K > 0$ such that $K(z) \geq c_K$ for all $|z| \leq \delta_K$. Moreover, it has compact support $[-1,1]$, $\frac{1}{2} <\int_{-1}^1 K^2(y)dy < \infty$ and the function is Lipschitz continuous, that is, $|K(v) - K(w)| \le C |v-w|$ for any $v,w \in \reals$ and some constant $C > 0$.
\end{enumerate} 
    
Assumption \ref{as:err} rules out cross-sectional dependence in the idiosyncratic errors assuming that the dependence structure across units is fully explained by the common covariates. This condition allows us to construct the multiscale test statistics by combining unit-wise Gaussian approximations without having to control for additional cross-sectional correlation. %Relaxing Assumption~(A5) would require explicit assumptions on cross-sectional dependence and substantially more involved technical arguments.
Assumption \ref{as:x-err1} imposes a homogeneous second-order structure for the idiosyncratic errors conditional on the common regressors $(\X_1,\dots,\X_T)$. Together with Assumption~\ref{as:err} and the representation in Assumption~\ref{as:x-err2}, this leads to a weakly dependent, second-order stationary structure for the processes $\{v_{it}\}_{t=1}^T$.

Assumption~\ref{as:x-err2} places the product process $\vv_{it}=\X_t\varepsilon_{it}$ in the physical-dependence framework of \cite{Wu2005}. This representation allows us to quantify temporal dependence via the physical dependence measures $\delta_q(g_i,t)$ and to invoke the strong approximation results of \cite{Karmakar2020}. Assumption \ref{as:x-err3} provides sufficient control of the tails of $\vv_{it}$.

Assumptions \ref{as:x-err4} and \ref{as:x-err5} correspond to the assumptions in \citet[Theorem 2.1-2.2]{Karmakar2020} that justify the strong approximation result. Assumption~\ref{as:x-err4} specifies how fast the physical dependence coefficients $\delta_q(g_i,t)$ must decay. It implies that the process $\{\vv_{it}\}_{t=1}^T$ is short-memory and yields a strong approximation error of order $T^{1/q}$. Note that one could allow for stronger serial dependence by choosing a smaller~$\xi$ at the price of a slower approximation rate. In that case, our main theoretical results would still remain valid. Assumption \ref{as:x-err5} is a common assumption in time series literature that ensures that the the eigenvalues of the covariance matrices of the increment processes are substantially large and away from zero, implying well-conditioned and positive definite covariance matrices.

Assumption~\ref{as:eig_xx} concerns the localized design matrix $\bm M_{XKX}(u, h)$ and requires this matrix to be positive definite with probability tending to one. This excludes local multicollinearity of the regressors on each interval $[u-h,u+h]$ in the grid $G_T$ and ensures that the kernel estimator $\hat\bfbeta_i(u,h)$ is well defined for all $(u,h)\in G_T$ and that the normalization by $\bm M_{XKX}(u,h)$ in the test statistic is stable.

Assumption \ref{as:GT_bound} governs the practical choice of the grid $\mathcal{G}_T$ and is a weak restriction commonly used in multiscale testing \citep{KhismatullinaVogt2020, KhismatullinaVogt2022}. The condition $|G_T|=O(T^\theta)$ for some fixed $\theta>0$ allows the number of location–bandwidth points to grow polynomially in $T$, which is large enough in practice to cover a rich collection of intervals. The lower bound $h^{\min}_T\gg T^{2/(3q)-1/3}$ prevents us from using bandwidths that are too small, that is, from forming intervals with too few observations for the limit theory to apply. The upper bound $h^{\max}_T\le 1/4$ rules out overly large bandwidths that would average over almost the entire sample and blur local features.

Assumption \ref{as:kernel} is a standard assumption in the nonparametric time series literature that is satisfied by many standard kernels such as the Epanechnikov kernel that we use in our simulation and application studies. The only non-standard restriction that we impose on the kernel function is the requirement $\frac{1}{2} <\int_{-1}^1 K^2(y)dy$ which is necessary for the anti-concentration bounds used in the Gaussian approximation of the multiscale statistic. This restriction is also satisfied by the most common kernel functions such as Epanechnikov (which we use in our empirical analysis), biweight or triweight kernels.

\subsection{Theoretical results}

In this section, we establish the theoretical properties of the multiscale test introduced in Section~\ref{sec:test}. We first show that the distribution of the test statistic $\widehat{\Psi}_T$ can be approximated by the distribution of a suitable Gaussian
max-type statistic $\Phi_T$. This strong approximation justifies the use of the simulated critical values $q_T(\alpha)$ defined in Section~\ref{sec:computation}. We then derive the asymptotic size of our test for the global null hypothesis $H_0$, study local power properties under a class of shrinking alternatives, and finally show that the procedure controls the family-wise error rate when simultaneously testing the full number of the local null hypotheses $H_0^{[i,j]}(u,h)$.

\begin{theorem} \label{thm:equiv_distr}
Let Assumptions \textcolor{blue}{\ref{as:trends} - \ref{as:kernel}} be fulfilled. Moreover, assume that the estimators $\widehat{\bm \Sigma}_i$ of the long-run covariance matrices $\bm \Sigma_i$ satisfy $\big|\big|\widehat{\bm \Sigma}_{i} - \bm \Sigma_{i}\big|\big|_{1} = o_p(\rho_T)$ for some $\rho_T = o(\log T)$, and that the long-run covariance matrices are
homogeneous across individuals, $\bm \Sigma_i = \bm \Sigma$ for all $i$. Then, under the global null $H_0$,
\begin{align*}
    \sup_{y\in \reals} \left|\pr(\widehat{\Psi}_{T} \leq y) - \pr(\Phi_{T} \leq y)\right| = o(1).
\end{align*}
\end{theorem}

Theorem~\ref{thm:equiv_distr} shows that, under $H_0$, the Kolmogorov distance between the distribution of the multiscale statistic $\widehat{\Psi}_T$ and that of the Gaussian test statistic $\Phi_T$ converges to zero as the sample size increases. The additional
assumption on $\widehat{\bm\Sigma}_{i}$ requires the long-run covariance matrices $\bm\Sigma_i$ to be estimated with sufficient accuracy. As discussed in Section~\ref{sec:lrv_estimation}, this rate can be achieved by standard HAC-type estimators. The homogeneity condition $\bm \Sigma_i = \bm \Sigma$ for all $i$ simplifies the covariance structure across units and allows us to work with a common Gaussian approximation. Intuitively, Theorem~\ref{thm:equiv_distr} ensures that the high-dimensional dependence across location–bandwidth points $(u,h) \in \mathcal{G}_T$ and pairs of units $(i,j)$ is adequately captured by the Gaussian coupling embodied in $\Phi_T$. Proof of Theorem \ref{thm:equiv_distr} is given in Appendix.

The result in Theorem \ref{thm:equiv_distr} is necessary for deriving the theoretical properties of the proposed multiscale test. Recall that $q_T(\alpha)$ denotes the $(1-\alpha)$-quantile of $\Phi_T$. First, we show that the test based on the critical value $q_T(\alpha)$ has asymptotically the correct size: the rejection probability under $H_0$ converges to the nominal level $\alpha$.

\begin{prop}[Asymptotic size] \label{prop:size}
Suppose that the assumptions of Theorem \ref{thm:equiv_distr} are fulfilled. Then, under $H_0$,
\begin{align*}
    \pr(\widehat{\Psi}_{T} \leq q_{T}(\alpha)) = (1-\alpha) + o(1).
\end{align*}
\end{prop}

Note that the size control is obtained despite the fact that the dimension of $\widehat{\Psi}_T$, that is, the number of location–bandwidth points $(u,h) \in G_T$ and ordered pairs $(i,j)$, may grow with $T$.

Second, we study the behavior of the test under a certain class of local alternatives and we show that under this certain class, our test has sufficient power. Let $\bfbeta_i = \bfbeta_{i,T}$ be a sequence of coefficient functions that depend on the time length $T$. The following class of alternatives ensures that, for some pair of units, sequences $\bfbeta_{i,T}$ and $\bfbeta_{j,T}$ are sufficiently separated on at least one interval in the multiscale grid.

\begin{prop}[Local power]\label{prop:power}
    Let the assumptions of Theorem \ref{thm:equiv_distr} be fulfilled. Moreover, assume that the following holds: for some pair $(i,j)$, there exists $(u,h)\in\mathcal{G}_T$ with $[u-h, u+h]\subseteq [0,1]$ and some coordinate $d\in\{1, \ldots, D\}$ such that \linebreak $\big|[\bfbeta_{i, T}(w) - \bfbeta_{j, T}(w)]_{d}\big| \geq c_T\sqrt{\log T/(Th)}$ for all $w\in[u-h, u+h]$, where $\{c_T\}$ is any sequence of positive numbers with $c_T\rightarrow\infty$ as $T\rightarrow\infty$. Then it holds that
    \begin{align*}
        \pr(\widehat{\Psi}_{T} \leq q_{T}(\alpha)) = o(1),
    \end{align*}
    and hence, the test rejects $H_0$ with probability tending to $1$.
\end{prop}

So far we have focused on the global null hypothesis $H_0$ of complete homogeneity across units. The construction of $\widehat{\Psi}_T$ also enables multiple testing of the collection of local null hypotheses $H_0^{[i,j]}(u,h)$ defined in Section~\ref{sec:hypothesis}. We now formalize the multiple testing problem and show that the procedure controls the family-wise error rate (FWER).

Let $\mathcal{M} = \{(u,h,i,j) : (u,h) \in \mathcal{G}_T \text{ and } 1\leq i<j\leq N\}$ denote the collection of all location-scale points $(u,h)\in \mathcal{G}_T$ and all pairs of units $(i,j)$. Let $\mathcal{M}_0 \subseteq \mathcal{M}$ denote the collection of tuples $(u,h,i,j)$ for which the local null hypothesis $H^{[i,j]}_0(u,h)$ is true. Our property of interest, FWER, is defined as the probability of rejecting $H^{[i,j]}_0(u,h)$ for at least one tuple $(u,h,i,j) \in \mathcal{M}_0$. That is,
\begin{align*}
    \mathrm{FWER}(\alpha) = \pr\left(\exists(u,h,i,j)\in \mathcal{M}_0: \hat{S}_{ij}(u,h) - \lambda(h) > q_T(\alpha)\right)
\end{align*}
for some given significance level $\alpha\in(0,1)$. We would like to bound the FWER, controlling the rate of false discoveries. The following proposition shows that our multiscale test asymptotically controls the FWER at level $\alpha$.
\begin{prop}\label{prop:fwer}
    Suppose that the assumptions of Theorem \ref{thm:equiv_distr} are fulfilled. Then, for any given $\alpha \in (0,1)$,
    \begin{align*}
        \mathrm{FWER}(\alpha) \leq \alpha + o(1).
    \end{align*}
\end{prop}
Equivalently, Proposition \ref{prop:fwer} implies that
\[
\mathbb{P}
\Bigl(
\forall \, (u,h,i,j) \in \mathcal{M}_0 :
\hat{S}_{ij}(u,h) - \lambda(h) \le q_T(\alpha)
\Bigr)
\;\ge\; 1 - \alpha + o(1),
\]
that is, with probability at least $1-\alpha$ (asymptotically), none of the true local null hypotheses is rejected by our procedure. This result provides the theoretical justification for the simultaneous confidence statement (R) in Section~3.4: the set of
accepted local hypotheses delivers a joint confidence statement about all coefficient functions across all units and all intervals, with an asymptotic coverage probability at least $1-\alpha$.

\section{Clustering}\label{sec:clust}
The global null hypothesis $H_0$ says that the coefficient functions are fully homogeneous across individuals. A rejection of $H_0$ implies the presence of heterogeneity in the coefficient functions. Rather than full heterogeneity across units, it is plausible that groups of individuals share the same coefficient functions - a structure commonly referred to as grouped heterogeneity \citep{BonhommeManresa2015, Su2016, Chen2019}. %\cite{Bonhomme2022} further argue that grouped heterogeneity serves as a useful approximation for more complex forms of heterogeneity in panel data models.
In this section, we discuss how our multiscale test can be used to uncover such group structure in the data. 

We assume that the group pattern follows the form
\begin{align*}
    \bfbeta_i = \sum^{K_0}_{k=1}\bfgamma_k\indic{i\in G_k},
\end{align*}

where $K_0$ is the number of groups, $G_k$ denotes the $k$-th group, with the partition $\bigcup_{k=1}^{K_0} G_k = \{1, 2, \ldots, N\}$ and $G_k \cap G_{k'} = \emptyset$ for $k \neq {k'}$. Each group shares a common coefficient function $\bfgamma_k$, and these functions differ across groups, i.e., $\bfgamma_k \neq \bfgamma_{k^\prime}$ for $k \neq k^\prime$. Since the group memberships are typically unknown, they need to be estimated from the data.

To estimate the group structure, we use partially aggregated individual multiscale test statistics $$\max_{(u,h)\in\mathcal{G}_T}\{\hat{S}_{ij}(u,h) - \lambda(h)\}$$ as a dissimilarity measure between series $i$ and $j$. Following \cite{Chen2019} and \cite{KhismatullinaVogt2022}, we use these distances as inputs for a hierarchical agglomerative clustering (HAC) algorithm. Given a specified number of groups $K_0$, the algorithm returns the estimated group memberships $\{\hat{G}_1, \ldots, \hat{G}_{K_0}\}$. We refer the reader to \cite{Chen2019}, \cite{KhismatullinaVogt2022}, and \cite{Hastie2009} for more details about the HAC in this context.

If the number of groups is unknown, we estimate it by selecting the smallest number of groups $\widehat{K}$ such that the maximum within-group distance does not exceed the critical value $q_T(\alpha)$.
%, that is,
%\begin{align*}
%    \widehat{K} = \min \{r=1,2\ldots| \text{maximum within-group distance} \leq q_{T}(\alpha)\},
%\end{align*}
%where ...
We show that $\widehat{K}$ is a consistent estimator of $K_0$ in the current setting.\footnote{Alternative selection procedures exist, such that the information criterion by \cite{Chen2019}, and may be validated in the current setting.} Below we state our assumptions on the group-specific coefficient functions.

\begin{enumerate}[label=(A\arabic*), leftmargin=1.05cm]
\setcounter{enumi}{12}
    \item \label{as:clusters} For each group $k= 1, \ldots, K_0$, the trend coefficient functions $\bfgamma_k = (\gamma_{k, 1}, \ldots, \gamma_{k, D})^\prime$ are Lipschitz continuous on $[0,1]$ coordinate-wise with normalization \linebreak $\int_0^1\gamma_{k, d} (u)du = 0$ for each $d = 1, \ldots, D$. Moreover, for any $k\neq k'$, the coefficient functions are different in the following way: There exists $(u,h) \in \mathcal{G}_T$ and a coordinate $d\in \{1, \ldots, D\}$ such that $
    \big|[\bfgamma_{k}(w) - \bfgamma_{k'}(w)]_d\big|_{\infty} \geq c_T\sqrt{\log T/(Th)}$ for all $w\in [u-h, u+h]$, where $0<c_T \rightarrow \infty$.
\end{enumerate} 

The first part of Assumption \ref{as:clusters} repeats Assumption \ref{as:trends}, whereas the second part is a necessary condition for the consistency of $\widehat{K}$ and is common in the literature.

The following proposition states that the hierarchical clustering algorithm produces consistent group-membership estimates and that the estimator $\widehat{K}$ of the true number of groups $K_0$ is consistent.

\begin{prop} \label{prop:cluster_fullT}
    Let the assumptions of Theorem \ref{thm:equiv_distr} and \ref{as:clusters} be fulfilled. Then we have that
    \begin{align*}
        \pr(\{\widehat{G}_1,\ldots, \widehat{G}_{\widehat{K}}\} = \{G_1,\ldots,G_{K_0}\}) \geq (1-\alpha) + o(1),
    \end{align*}
    and 
    \begin{align*}
        \pr(\widehat{K}=K_0)\geq (1-\alpha) + o(1).
    \end{align*}
\end{prop}
Importantly, our multiscale approach allows us to identify the locations where two group-specific coefficients $\bfgamma_k$ and $\bfgamma_{k'}$ are different from each other. For each pair of groups $(k, k')$, let $\mathcal{A}_{T}^{[k, k']}$ denote the set of $(u,h)\in\mathcal{G}_T$ for which there is at least one local difference between individuals across groups, that is,
\begin{align*}
    \mathcal{A}^{[k, k']}_T(\alpha) = \{(u,h)\in \mathcal{G}_T: \hat{S}_{ij}(u,h) > q_T(\alpha) \text{ for some }i\in\widehat{G}_{k}, j\in \widehat{G}_{k'}\}.
\end{align*}
We define the set $\mathcal{S}_T^{[k, k']}(\alpha)$ as the set of intervals $[u-h, u+h] \subseteq [0, 1]$ for which $(u,h)\in\mathcal{A}^{[k, k']}_T(\alpha)$, i.e.,
\begin{align*}
    \mathcal{S}_T^{[k, k']}(\alpha) = \{[u-h, u+h]: (u,h)\in \mathcal{A}^{[k, k']}_T(\alpha)\}.
\end{align*}
Moreover, let 
\begin{align*}
    E_T^{[k, k']}(\alpha) = \{\forall [u-h, u+h] \in \mathcal{S}^{[k, k']}_T(\alpha): \bfgamma_{k}(v) \neq \bfgamma_{k'}(v) \text{ for some }v\in[u-h, u+h]\}
\end{align*}
be the event that the group-specific coefficients differ on all intervals in $ \mathcal{S}^{[k, k']}_{T}(\alpha)$. The following proposition says the multiscale test correctly identifies the time intervals where the group-specific coefficients $\bfgamma_{k}$ and $\bfgamma_{k'}$ are different from each other with asymptotic confidence $1-\alpha$.
%MAKE IT CONSISTENT WITH THE ASSUMPTION A13

\begin{prop} \label{prop:cluster_groupseperation}
    Let the assumptions of Proposition \ref{prop:cluster_fullT} be fulfilled. Then we have that the event 
    \begin{align*}
        E_T(\alpha) = \{\cap_{1\leq k<k'\leq\hat{K}}E^{[k, k']}_T(\alpha)\} \cap \{\widehat{K} = K_0 \text{ and }\widehat{G}_{k} = G_{k} \text{ for all }k \}
    \end{align*}
    asymptotically occurs with probability at least $1-\alpha$, that is,
    \begin{align*}
        \pr\left(E_T(\alpha)\right) \geq (1-\alpha) + o(1).
    \end{align*}
\end{prop}

\begin{remark}
In this work we assume that the group structure is the same across all $D$ coordinates. Recent works by \cite{Cheng2019} and \cite{Leng2023} discuss models with multi-dimensional group structure, where different parameters may correspond to different group structures. We leave this extension for further research.
\end{remark}

%\subsection{Common regressors}
%One can allow for common regressors, that is, $\X_{it}=\X_t$ for all $i$.
%\input{Sections/Simulations}
\section{Empirical application} \label{sec:application}
In our empirical study, we revisit the relationship between U.S. interest rates and foreign output. \cite{Iacoviello2019} find that U.S. monetary tightening results in larger drops in GDP for emerging economies than for advanced economies, and they document substantial heterogeneity in the effect of U.S. monetary shocks on foreign economies. We employ our multiscale test to test for heterogeneity across economies and across time. While doing so, we allow for a time-varying relationship between U.S. monetary shocks and foreign GDP, as opposed to the static relationship in \cite{Iacoviello2019}. 

The quarterly interest rate and foreign GDP data spans the period between 1965-Q1 and 2016-Q2, and covers 49 foreign countries. We include the U.S. GDP in our GDP data, bringing the number of time series to $N=50$. The dataset is taken from \cite{Iacoviello2019} and is publicly available at the authors' website.\footnote{\url{https://www.matteoiacoviello.com/research.htm}} For the detailed description of the dataset we refer the reader to the original paper.

For our analysis, we construct the U.S. monetary shock variable (denoted further by $X_t$) as a residual of a regression on several control variables as described by \cite{Iacoviello2019}. The GDP data (denoted by $y_{it}$ for a country $i$) is de-seasonalized by extracting the residual from an country-specific regression on the fourth-order lag and a quadratic trend (corresponding to the controls in \cite{Iacoviello2019}). Afterwards, we demean each of the obtained time series separately in order to account for the individual fixed-effects. As a result, we estimate the time-varying  GDP response to an U.S. monetary shock from the following model: 
\begin{align*}
    y_{i(t+h)} = \beta_{i}(t/T)X_t + \epsilon_{it},
\end{align*}
for some lag $h$. In what follows, we fix $h=1$ in the current draft.

%We perform the test exactly as described in Section \ref{sec:simulations}. We find that 
The test is applied exactly as described in Section \ref{sec:computation}. The nonparametric estimators are computed as described in Section \ref{sec:estim}, where we employ the Epanechnikov kernel. We estimate the long-run variances $\bm \Sigma_i$, $i=1,\ldots,N$, by the Bartlett-kernel HAC estimator with data-driven bandwidth (see Section \ref{sec:lrv_estimation}). We construct the grid $\mathcal{G}_T$ following \cite{KhismatullinaVogt2022} as $\mathcal{G}_T = U_T \times H_T$ with
\begin{align*}
    &U_T = \{u\in[0,1]: u=\frac{5t}{T} \text{ for some } t\in\naturals\},\\
    &H_T = \left\{h\in\left[\frac{1}{T^{1/3}}, \frac{1}{4}\right]: h = \frac{5t-3}{T} \text{ for some } t\in\naturals\right\}.
\end{align*}

\begin{figure}
    \centering
    \includegraphics[width=0.8\linewidth]{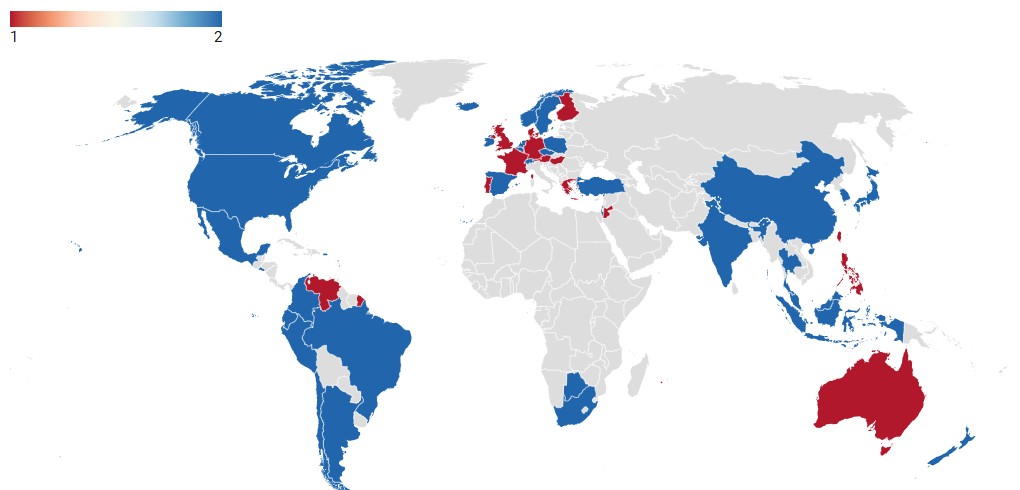}
    \caption{Estimated cluster memberships using hierarchical clustering. The number of clusters is set to $K=2$.}
    \label{fig:app_clusters}
\end{figure}

Furthermore, we compute the critical value $q_T(\alpha)$ of our test by Monte Carlo simulation based on $B=1000$ Gaussian samples. The results are as follows. First, we reject the null hypothesis of global homogeneity at $5\%$ level. This indicates that there is a heterogeneous effect of U.S. monetary shocks on foreign GDP. To investigate the heterogeneity further, we partition the countries into clusters. Given the natural partition into advanced vs. emerging countries and to be able to directly compare with \cite{Iacoviello2019}, we use $K=2$ clusters.\footnote{Our selection procedure indicates $K=18$ clusters but for interpretability reasons we chose to report the results for $K=2$ clusters.} We visualize the clusters for $K=2$ in Figure \ref{fig:app_clusters}. The figure illustrates that the second cluster is mostly made up from European countries, while the first clusters covers South America and Asia. The clusters correspond mostly with the partition into advanced and emerging economies made by \cite{Iacoviello2019}, although some differences arise. We find that the time-varying effect of the U.S. monetary shock are similar for the U.S. and Canadian economies, however, we find significant differences between these economies and most European countries which together with Australia mostly constitute a separate cluster. Our results indicate that allowing for time-varying exposure to U.S. monetary shocks, the heterogeneity in exposure may be more sophisticated than the classical advanced vs. emerging economies partition that is currently used. To sum up, our test serves as a statistically rigorous starting point for further analysis of the various economic relationships.
%\newpage
\section{Conclusion} \label{sec:conclusion}

In this paper, we introduce a multiscale test for heterogeneous time-varying coefficients in panels and prove that it asymptotically controls the family-wise error rate. We also show that under local alternatives our test has asymptotic power of one. The proposed  multiscale procedure allows us to make simultaneous confidence statements about which of the coefficient functions are different and where these differences occur. Furthermore, the results of our test can be used as a dissimilarity measure for a subsequent HAC clustering algorithm. %Simulations demonstrate correct size and high power across realistic designs, while 
For the real-life application, we study the relationship between U.S interest rates and foreign output of $49$ economies and the economy of U.S. itself. The results show that foreign GDP responses to U.S. monetary shocks exhibit heterogeneity and can be partitioned into two clusters with crisis-specific dynamics. %The methodology is implemented in the open-source GitHub repository along with replication material, providing an easy start for applying our method to other economic problems. 
Future research will relax the common-regressor assumption and extend strong approximation theory to other forms of dependencies.
%\newpage
\section{Acknowledgement}
The authors thank Chen Zhou and Wendun Wang for their insightful suggestions during the preparation of the paper.

\newpage
\begin{spacing}{1.6}
\bibliographystyle{ims}
\bibliography{bib}

\begin{thebibliography}{4}
\expandafter\ifx\csname natexlab\endcsname\relax\def\natexlab#1{#1}\fi
\expandafter\ifx\csname url\endcsname\relax
  \def\url#1{\texttt{#1}}\fi
\expandafter\ifx\csname urlprefix\endcsname\relax\def\urlprefix{URL }\fi
\providecommand{\eprint}[2][]{\url{#2}}

\bibitem[{Chernozhukov et~al.(2017)Chernozhukov, Chetverikov and Kato}]{Chernozhukov2017}
\textsc{Chernozhukov, V.}, \textsc{Chetverikov, D.} and \textsc{Kato, K.} (2017).
\newblock Central limit theorems and bootstrap in high dimensions.
\newblock \textit{Annals of Probability}, \textbf{45} 2309--2352.

\bibitem[{Karmakar and Wu(2020)}]{Karmakar2020}
\textsc{Karmakar, S.} and \textsc{Wu, W.~B.} (2020).
\newblock Optimal {G}aussian approximation for multiple time series.
\newblock \textit{Statistica Sinica}, \textbf{30} 1399--1417.

\bibitem[{Khismatullina and Vogt(2022)}]{KhismatullinaVogt2022}
\textsc{Khismatullina, M.} and \textsc{Vogt, M.} (2022).
\newblock Multiscale comparison of nonparametric trend curves.
\newblock \textit{arXiv preprint arXiv:2209.10841}.

\bibitem[{Nazarov(2003)}]{Nazarov2003}
\textsc{Nazarov, F.} (2003).
\newblock On the maximal perimeter of a convex set in $\reals^n$ with respect to a {G}aussian measure.
\newblock In \textit{Geometric Aspects of Functional Analysis}. Springer, 169--187.

\end{thebibliography}
\end{spacing}

% \newpage
% \setcounter{page}{1}
% \renewcommand{\thepage}{S\arabic{page}}
% \centerline{\LARGE Supplemental Appendix to:}
% \bigskip
% \centerline{\LARGE Multiscale comparison of nonparametric trending coefficients}
% \bigskip
% \centerline{\large \textit{Marina Khismatullina$^{1}$, Bernhard van der Sluis$^{1}$}}
% \bigskip
% \centerline{$^1$\ \textit{Econometric Institute, Erasmus University Rotterdam}}
% \bigskip
% \noindent{\Large\textbf{Contents}}\\[2ex]
% Appendix~\ref{sec:proofs}:\ \nameref{sec:proofs}\hfill\pageref{sec:proofs}\\[1ex]
% Appendix~\ref{sec:additional_proofs}:\ \nameref{sec:additional_proofs}\hfill\pageref{sec:additional_proofs}\\[1ex]
%%
\clearpage
\appendix
\begingroup
\begin{spacing}{1.5} 
\small
\renewcommand{\thesection}{A}
\section{Proofs}\label{sec:proofs}
\renewcommand{\thedefinition}{\thesection.\arabic{definition}}
\renewcommand{\theequation}{\thesection.\arabic{equation}}
\renewcommand{\thelemma}{\thesection.\arabic{lemma}}
\renewcommand{\theprop}{\thesection.\arabic{prop}}
\renewcommand{\thetheorem}{\thesection.\arabic{theorem}}
\setcounter{equation}{0}
\setcounter{theorem}{0}
%\newcounter{lemma} \newcounter{prop}
%\setcounter{lemma}{0}
%\setcounter{prop}{0}

\begin{proof}[\textnormal{\textbf{Proof of Theorem \ref{thm:equiv_distr}}}] With the help of this theorem, we justify that inference using the critical values from Section \ref{sec:test} is valid. We first summarize the main proof strategy, which splits up into three steps, and then fill in the details. We will also make use of an auxiliary test statistics 
\begin{align*}
\widehat{\Psi}^0_{T} = \max_{1 \le i < j \le N} \max_{(u,h)\in \mathcal{G}_T} \big\{ \hat{S}^0_{ij}(u,h) - \lambda(h) \big\},
\end{align*}
where
\begin{align*}
    &\hat{S}^0_{ij}(u,h) = \Bigg|\Bigg|\widehat{\bm\Sigma}^{-1/2}_{ij}\left(\frac{1}{\sqrt{Th}}\sum_{t=1}^T \X_{t} \left(\left(\varepsilon_{it} - \bar{\varepsilon}_i\right) - \left(\varepsilon_{jt} - \bar{\varepsilon}_j \right)\right) K_{t, u, h}\right)\Bigg|\Bigg|_{\infty},
\end{align*}
as specified in \eqref{eq:individ_under_null}. $\hat{S}^0_{ij}(u,h)$ is different from $\hat{S}_{ij}(u,h)$ only in the fact that it does not depend on the coefficient functions $\bfbeta_i$ and $\bfbeta_j$. Under the null $H_0$, when we assume that $\bfbeta_i \equiv \bfbeta_j$ for all $1\leq i < j \leq N$, $\widehat{\Psi}_{T}$ is exactly equal to $\widehat{\Psi}^0_{T}$.

\subsubsection*{Step 1}

%To start with, we consider a simplified setting where we assume that for all $i=1, \ldots, N$ we set $\hat{\beta}^*_{i, h_0}(u)$ to zero everywhere on $[0, 1]$ at the first step, which leads to the simpler statistic 
% \[ \doublehattwo{\Psi}_{N,T} = \max_{1 \le i < j \le n} \max_{(u,h) \in \mathcal{G}_T}\Big\{\doublehat{S}^0_{ij}(u, h) - \lambda(h)\Big\}, \, \text{ with } \]
% {\footnotesize
% \begin{align*}
%     \doublehat{S}^0_{ij}(u, h) = \sup_{x\in \mathcal{X}}\frac{1}{\sqrt{\big(\doublehat{A}^{i}_{T,u,h}(x) + \doublehat{A}^{j}_{T,u, h}(x)\big) Q^{-2}_{T,u, h}(x)}} \frac{\left|\frac{1}{T}\sum_{t=1}^T X_{t} \Big(\big(\varepsilon_{it} - \bar{\varepsilon}_i\big) - \big(\varepsilon_{jt} - \bar{\varepsilon}_j\big)\Big) K\left(\frac{t/T - u}{h}\right)\right|}{\frac{1}{T}\sum_{t=1}^T X^2_{t} K\left(\frac{t/T - u}{h}\right)}.
% \end{align*}}

% and $\doublehat{A}^{i}_{T,u, h}(x)$ is computed in exactly the same way as $\widehat{A}^{i}_{T,u, h}(x)$ except that all occurrences of $\hat{\beta}^*_{i, h_0}$ are replaced by $0$. By assumptions \textcolor{blue}{(check this!)}, $\widehat{A}^i_{T,u, h}(x) = A_{T,u, h}(x) + o_{pu}(\rho_T)$ with $\rho_T = \ldots$. In what follows, we take for granted that the estimator $\doublehat{A}^{i}_{T,u, h}(x)$ has this property as well.

We first have a closer look at the statistic $\widehat{\Psi}^0_{T}$. In particular, we show that there exists an identically distributed version $\widetilde{\Psi}_{T}$ of $\widehat{\Psi}^0_{T}$ which is close to the Gaussian statistic $\Phi_{T}$ from \eqref{eq:gauss_stat}. More formally, we prove the following result.  
\begin{prop}\label{prop:strong_approx}
There exist statistics $\{ \widetilde{\Psi}_{T}: T =1,2,\ldots \}$ with the following two properties: (i) $\widetilde{\Psi}_{T}$ has the same distribution as $\widehat{\Psi}^0_{T}$ for any $T$, and (ii)
\begin{equation*}
\big| \widetilde{\Psi}_{T} - \Phi_{T} \big| = o_p(\delta_T),
\end{equation*}
where $\delta_T = T^{1/q}/\sqrt{T(h^{\mathrm{min}}_T)^3} + \rho_T\sqrt{\log T}$ and $\Phi_{T}$ is a Gaussian statistic as defined in \eqref{eq:gauss_stat}. 
\end{prop}
The proof makes heavy use of strong approximation theory for dependent processes. As it is quite technical, we defer the full proof to the next subsection.

\subsubsection*{Step 2}
In this step, we establish some properties of the Gaussian statistic $\Phi_{T}$. Specifically, we prove the following result.  
\begin{prop}\label{prop:anticon}
It holds that 
\begin{equation*}
\sup_{y \in \reals} \pr \big( | \Phi_{T} - y | \le \delta_T \big) = o(1),
\end{equation*}
where $\delta_T = \frac{T^{1/q}}{\sqrt{T (h^{\mathrm{min}}_T})^3} + \rho_T\sqrt{\log T}$.
\end{prop}
Roughly speaking, this proposition says that the random variable $\Phi_{T}$ does not concentrate too strongly in small regions of the form $[y-\delta_T, y+\delta_T]$ with $\delta_T$ converging to $0$. The main technical tool for deriving it are anti-concentration bounds for Gaussian random vectors. The proof of this proposition with $\delta_T = T^{1/q}/\sqrt{T(h^{\mathrm{min}}_T})^3 + \rho_T\sqrt{\log T}$ can be found later in this section and closely follows the proof in \citeapndx{KhismatullinaVogt2022} (Proposition A.8).

\subsubsection*{Step 3}
We now use Steps 1 and 2 to prove that 
\begin{equation*}\label{eq:claim-step3}
\sup_{y \in \reals} \big| \pr(\widehat{\Psi}^0_{T} \le y) - \pr(\Phi_{T} \le y ) \big| = o_P(1). 
\end{equation*}

The proof of this proposition can be found in \citeapndx{KhismatullinaVogt2022}, statement (A.5).
\end{proof}

\begin{proof}[\textnormal{\textbf{Proof of Proposition \ref{prop:cluster_fullT}}}]
     The proof of this results follows from the steps in the proof of Proposition 5.1 in \cite{KhismatullinaVogt2022} and therefore omitted for brevity.
\end{proof}

\begin{proof}[\textnormal{\textbf{Proof of Proposition \ref{prop:cluster_groupseperation}}}]
We define the auxiliary statistic
\begin{align}
    \hat{\Psi}^0_{T} = \max_{1\leq i<j\leq N}\max_{(u,h)\in \mathcal{G}_T}\{\hat{S}^0_{ij}(u,h) - \lambda(h)\},
\end{align}
where $\hat{S}^0_{ij}(u,h)$ is defined in \eqref{eq:individ_under_null}. Let us consider the event
    \begin{align*}
        B_{T} = \{\hat{\Psi}^0_{T} \leq q_T(\alpha) \quad \text{and} \quad \min_{1\leq k < k'\leq K_0} \min_{\substack{i\in G_k\\ j \in G_{k'}}} \hat{\Psi}_{ij,T} \geq q_T(\alpha)\},
    \end{align*}
    where the statistic $\hat{\Psi}_{ij,T}$ is defined as
    \begin{align*}
       \hat{\Psi}_{ij,T} = \max_{(u,h) \in \mathcal{G}_T} \{\hat{S}_{ij}(u,h) -\lambda(h)\},
    \end{align*}
    for each $i$ and $j$. In other words, $\min_{1\leq k < k'\leq K_0} \min_{\substack{i\in G_k\\ j \in G_{k'}}} \hat{\Psi}_{ij,T}$ corresponds to the smallest between-group multiscale distance. In case event $B_T$ holds, it holds that
    \begin{align*}
        \hat{\Psi}^0_{T} < \min_{1\leq k < k'\leq K_0} \min_{\substack{i\in G_k\\ j \in G_{k'}}} \hat{\Psi}_{ij,T}.
    \end{align*}
    Since $\hat{\Psi}^0_T$ is identical to $\hat{\Psi}_T$ under the null $H_0$, we can use Proposition \ref{prop:size} to conclude that
    \begin{align*}
        \pr\left(\hat{\Psi}^0_T \leq q_T(\alpha)\right) \geq (1-\alpha) + o(1).
    \end{align*}
    Furthermore, applying the same arguments used for Proposition \ref{prop:power} yields that
    \begin{align*}
        \pr\left(\min_{1\leq k < k'\leq K_0} \min_{\substack{i\in G_k\\ j \in G_{k'}}} \hat{\Psi}_{ij,T} \leq q_T(\alpha)\right).
    \end{align*}
    Therefore, we obtain that
    \begin{align} \label{eq:B_power}
        \pr\left(B_T\right) \geq (1-\alpha) + o(1).
    \end{align}
    Analogous to the proof of Proposition 5.1 in \cite{KhismatullinaVogt2022}, we can show that
    \begin{align}\label{eq:BsubK}
        B_T \subseteq \{\hat{K} = K_0 \text{ and } \hat{G}_k = G_k \text{ for all }k\},
    \end{align}
    as a result of the HAC algorithm to obtain clusters. Moreover, since for all $i,j$ and $(u,h) \in \mathcal{G}_T$ for which it holds that
    \begin{align*}
        \hat{S}^0_{ij}(u,h) - \lambda(h) \geq q_T(\alpha) \quad \text{and} \quad \hat{S}_{ij}(u,h) - \lambda(h) \geq q_T(\alpha),
    \end{align*}
    it must hold that 
    \begin{align*}
    \sum^T_{t=1}\X_t\bigg(\X_{t}^\prime\bfbeta_i(t/T) - \X_{t}^\prime\bfbeta_j(t/T)\bigg)K_{t,u,h} \neq 0, 
    \end{align*}
    which implies that $\bfbeta_i(v) - \bfbeta_j(v) \neq 0$ for some $v\in [u-h, u+h]$. Therefore, we obtain that
    \begin{align} \label{eq:BsubE}
        B_T \subseteq \bigcap_{1\leq k < k'\leq \hat{K}}E_T^{k,k'}(\alpha). 
    \end{align}
    Combining the results in \eqref{eq:BsubK} and \eqref{eq:BsubE}, we obtain that
    \begin{align*}
        B_T \subseteq \Big\{\bigcap_{1\leq k < k'\leq \hat{K}}E_T^{k,k'}(\alpha)\Big\} \cap \{\hat{K} = K_0 \text{ and } \hat{G}_k = G_k \text{ for all }k\} = E_T(\alpha).
    \end{align*}
    Hence, result \eqref{eq:B_power} implies that $\pr\left(E_T(\alpha)\right) \geq (1-\alpha) + o(1)$.
\end{proof}
%\newpage
\renewcommand{\thesection}{B}
\section{Additional Proofs}\label{sec:additional_proofs}
\renewcommand{\thedefinition}{\thesection.\arabic{definition}}
\renewcommand{\theequation}{\thesection.\arabic{equation}}
\renewcommand{\thelemma}{\thesection.\arabic{lemma}}
\renewcommand{\theprop}{\thesection.\arabic{prop}}
\renewcommand{\thetheorem}{\thesection.\arabic{theorem}}
\setcounter{equation}{0}
\setcounter{theorem}{0}

\begin{proof}[Proof of Proposition \ref{prop:strong_approx}]
    Consider the stationary $D$-dimensional process $\mathcal{N}_i = \{\vv_{it} = \X_{t}\varepsilon_{it}: 1 \leq t \leq T\}$ for some fixed $i \in \{1,\ldots,N\}$. By Theorem 2.2 in \citeapndx{Karmakar2020}, the following strong approximation result holds true: On a richer probability space, there exist a sequence of random vectors $\widetilde{\mathcal{N}}_i = \{ \widetilde{\vv}_{it}: t \in \naturals \}$ %and a sequence of a mean zero independent Gaussian vectors $\{\Z_{it}: t\in \naturals\}$
    such that $[\widetilde{\vv}_{i1},\ldots,\widetilde{\vv}_{iT}] \stackrel{\mathcal{D}}{=} [\vv_{i1},\ldots,\vv_{iT}]$ for each $T$ and 
    \begin{align} \label{eq-strongapprox-dep}
    \max_{1\leq t\leq T} \Bigg|\Bigg|\sum^t_{s=1}\widetilde{\vv}_{is} - \bm \Sigma_i^{1/2}IB(t)\Bigg|\Bigg| = o_p(T^{1/q}),
    \end{align}
where $IB(\cdot)$ is a centered Brownian motion in $\reals^D$ and $\bm \Sigma_i = \sum_{k\in\integers} Cov(\vv_{i0}, \vv_{ik})$ is the long-run covariance matrix of $\{\vv_{it}: 1 \leq t \leq T\}$. Let $\widetilde{S}_{ij}(u, h)$  be defined similarly to $\widehat{S}^0_{ij}(u, h)$ but with $\widetilde{\vv}_{it}$ instead of $\vv_{it}$:
\[\widetilde{S}_{ij}(u, h) = \Bigg|\Bigg|\widetilde{\bm \Sigma}^{-1/2}_{ij}\left(\frac{1}{\sqrt{Th}}\sum_{t=1}^T \Big( \big(\widetilde{\vv}_{it} - \bar{\widetilde{\vv}}_i\big) - \big(\widetilde{\vv}_{jt} - \bar{\widetilde{\vv}}_j \big)\Big) K_{t, u, h}\right)\Bigg|\Bigg|_{\infty},\]

where the estimator $\widetilde{\bm \Sigma}_{ij}$ is constructed from the samples $\widetilde{\mathcal{N}}_i$ and $\widetilde{\mathcal{N}}_j$ in the same way as $\widehat{\bm \Sigma}_{ij}$ is constructed from $\mathcal{N}_i$ and $\mathcal{N}_j$. Note that since $[\widetilde{\vv}_{i1},\ldots,\widetilde{\vv}_{iT}] \stackrel{\mathcal{D}}{=} [\vv_{i1},\ldots,\vv_{iT}]$ and $\big|\big|\widehat{\bm \Sigma}^{-1/2}_{ij} - \bm \Sigma_{ij}^{-1/2}\big|\big|_{1}=o_p(\rho_T)$, we have that $\big|\big|\widetilde{\bm \Sigma}_{ij}^{-1/2} - \bm \Sigma_{ij}^{-1/2}\big|\big|_{1} = o_p(\rho_T)$ (which we will use later in the proof).

Similarly to $\widehat{\Psi}^0_T$, $\widetilde{\Psi}_{T}$ is defined as 
\[ \widetilde{\Psi}_{T} = \max_{1 \le i < j \le N}\max_{(u,h) \in \mathcal{G}_T}  \left\{ \widetilde{S}_{ij}(u, h) - \lambda(h)\right\}.\]
Moreover, we introduce an auxiliary statistic 
\begin{align*}
\Phi^{\diamond}_{T} = \max_{1 \le i < j \le N}\max_{(u,h) \in \mathcal{G}_T} \left\{S^{\diamond}_{ij}(u, h) - \lambda(h)\right\},
\end{align*}
where 
\[ S^{\diamond}_{ij}(u, h) =  \Bigg|\Bigg|\widetilde{\bm \Sigma}^{-1/2}_{ij}\left(\frac{1}{\sqrt{Th}}\sum_{t=1}^T\Big( \bm \Sigma^{1/2}_i\big(\Z_{it} - \bar{\Z}_i\big) - \bm \Sigma^{1/2}_j\big(\Z_{jt} - \bar{\Z}_j \big)\Big) K_{t, u, h}\right)\Bigg|\Bigg|_{\infty},\]

which is different from $S_{ij}(u, h)$ only in using $\widetilde{\bm \Sigma}^{-1/2}_{ij}$ instead of $\bm \Sigma^{-1/2}_{ij}$ as a normalization matrix. As a reminder, we also state here the formula for $\Phi_{T}$:
\begin{align*}
\Phi_{T} = \max_{1 \le i < j \le N}\max_{(u,h) \in \mathcal{G}_T} \left\{S_{ij}(u, h) - \lambda(h)\right\},
\end{align*}
with 
\[ S_{ij}(u, h) =  \Bigg|\Bigg|\bm \Sigma^{-1/2}_{ij}\left(\frac{1}{\sqrt{Th}}\sum_{t=1}^T\Big( \bm \Sigma^{1/2}_i\big(\Z_{it} - \bar{\Z}_i\big) - \bm \Sigma^{1/2}_j\big(\Z_{jt} - \bar{\Z}_j \big)\Big) K_{t, u, h}\right)\Bigg|\Bigg|_{\infty}.\]
Using this notation, we trivially have 
\begin{align*}
    \Big|\widetilde{\Psi}_{T} - \Phi_{T}\Big| \leq \Big|\widetilde{\Psi}_{T} - \Phi^{\diamond}_{T}\Big| + \Big|\Phi^{\diamond}_{T} - \Phi_{T}\Big|.
\end{align*}
In what follows, we prove that 
\begin{align}
\big| \widetilde{\Psi}_{T} - \Phi^{\diamond}_{T} \big| & = o_p \left( \frac{T^{1/q}}{\sqrt{T(h^{\min}_T)^3}} \right) \label{eqA:psitilde-phidiamond},\\
\big|\Phi^{\diamond}_{T} - \Phi_{T} \big| & =o_p\Big(\rho_T\sqrt{\log T}\Big). \label{eqA:phidiamond-phi}
\end{align}
For the sake of simplicity, we introduce the following notation:
\begin{align*}
    \veca_{ij, T}(u, h) &= \frac{1}{\sqrt{Th}}\sum_{t=1}^T \big\{ (\widetilde{\vv}_{it} - \bar{\widetilde{\vv}}_i)  - (\widetilde{\vv}_{jt} - \bar{\widetilde{\vv}}_j)\big\}K_{t, u, h}\\
    \vecb_{ij, T}(u, h) &= \frac{1}{\sqrt{Th}}\sum_{t=1}^T  \{ \bm \Sigma^{1/2}_i(\Z_{it} - \bar{\Z}_i) - \bm \Sigma^{1/2}_j(\Z_{jt} - \bar{\Z}_j)\}K_{t, u, h}.
\end{align*}
We then have that 
\begin{align*}
    \widetilde{S}_{ij, T}(u, h)  &=  \big|\big|\widetilde{\bm \Sigma}^{-1/2}_{ij}\veca_{ij, T}(u, h)\big|\big|_{\infty},\\
    S^{\diamond}_{ij, T}(u, h) &=  \big|\big|\widetilde{\bm \Sigma}^{-1/2}_{ij}\vecb_{ij, T}(u, h)\big|\big|_{\infty},\\
    S_{ij, T}(u, h)  &=  \big|\big|\bm \Sigma^{-1/2}_{ij}\vecb_{ij, T}(u, h)\big|\big|_{\infty}.
\end{align*}
First, we consider $\big| \widetilde{\Psi}_{T} - \Phi^{\diamond}_{T} \big|$. Due to the triangle inequality, 
\begin{align*}
 \big| \widetilde{S}_{ij, T}(u,h) - S^{\diamond}_{ij, T}(u,h) \big| &= \Bigg| \Big|\Big|\widetilde{\bm \Sigma}^{-1/2}_{ij}\veca_{ij, T}(u, h)\Big|\Big|_{\infty} - \Big|\Big|\widetilde{\bm \Sigma}^{-1/2}_{ij}\vecb_{ij, T}(u, h)\Big|\Big|_{\infty}\Bigg|\\
 & \leq \Big|\Big|\widetilde{\bm \Sigma}^{-1/2}_{ij}\big(\veca_{ij, T}(u, h) - \vecb_{ij, T}(u, h)\big)\Big|\Big|_{\infty}.
\end{align*}

Now using the definition of matrix-vector multiplication, we get the following
\begin{align*}
    &\Big|\Big|\widetilde{\bm \Sigma}^{-1/2}_{ij}\big(\veca_{ij, T}(u, h) - \vecb_{ij, T}(u, h)\big)\Big|\Big|_{\infty} \\
    &\quad= \max_{d = 1, \ldots, D} \bigg| \Big[\widetilde{\bm \Sigma}^{-1/2}_{ij}\big(\veca_{ij, T}(u, h) - \vecb_{ij, T}(u, h)\big)\Big]_d\bigg|\\
    &\quad=\max_{d =1, \ldots, D} \bigg| \sum_{e=1}^D\Big[\widetilde{\bm \Sigma}^{-1/2}_{ij}\Big]_{e,d}\Big[\veca_{ij, T}(u, h) - \vecb_{ij, T}(u, h)\Big]_e\bigg|\\
    &\quad\leq \max_{d =1, \ldots, D}  \sum_{e=1}^D\bigg|\Big[\widetilde{\bm \Sigma}^{-1/2}_{ij}\Big]_{e,d}\Big[\veca_{ij, T}(u, h) - \vecb_{ij, T}(u, h)\Big]_e\bigg|\\
    &\quad\leq\max_{e=1, \ldots, D}\bigg|\Big[\veca_{ij, T}(u, h) - \vecb_{ij, T}(u, h)\Big]_e\bigg| \cdot \max_{d =1, \ldots, D}  \sum_{e=1}^D\bigg|\Big[\widetilde{\bm \Sigma}^{-1/2}_{ij}\Big]_{e,d}\bigg| \\
    &\quad=\big|\big|\veca_{ij, T}(u, h) - \vecb_{ij, T}(u, h)\big|\big|_{\infty}  \cdot\big|\big|\widetilde{\bm \Sigma}^{-1/2}_{ij}\big|\big|_{1}.
\end{align*}
Then,
\begin{equation}
\begin{aligned}
\big| \widetilde{\Psi}_{T} - \Phi^{\diamond}_{T}   \big| 
 & \le\max_{1\le i < j \le N} \max_{(u,h) \in \mathcal{G}_T} \big| \widetilde{S}_{ij, T}(u,h) - S^{\diamond}_{ij, T}(u,h) \big|\label{eqA:psitilde-phidiamond:1}\\
 & \le \max_{1\le i < j \le N} \big|\big|\widetilde{\bm \Sigma}^{-1/2}_{ij}\big|\big|_{1} \max_{1\le i < j \le N} \max_{(u,h) \in \mathcal{G}_T} \big|\big|\veca_{ij, T}(u, h) - \vecb_{ij, T}(u, h)\big|\big|_{\infty},
\end{aligned}
\end{equation}
Now, using summation by parts we get that
\begin{align*}
    &\veca_{ij, T}(u, h)-\vecb_{ij, T}(u, h) \\
    &= \frac{1}{\sqrt{Th}}\sum_{t=1}^T K_{t, u, h} \big\{ (\widetilde{\vv}_{it} - \bar{\widetilde{\vv}}_i) - (\widetilde{\vv}_{jt} - \bar{\widetilde{\vv}}_j) - \bm \Sigma^{1/2}_i(\Z_{it} - \bar{\Z}_i) + \bm \Sigma^{1/2}_j(\Z_{jt} - \bar{\Z}_j) \big\}  \\
 & = \frac{1}{\sqrt{Th}}\sum_{t=1}^{T-1} \vecc_{ij, t} \left(K_{t, u, h} -K_{t+1, u, h}\right) + \frac{1}{\sqrt{Th}}\vecc_{ij, T} K_{T, u, h},
\end{align*}
where 
\begin{align*}
\vecc_{ij, t} = \sum_{s=1}^t \big\{ (\widetilde{\vv}_{is} - \bar{\widetilde{\vv}}_i)  - (\widetilde{\vv}_{js} - \bar{\widetilde{\vv}}_j) - \bm \Sigma^{1/2}_i(\Z_{is} - \bar{\Z}_i) + \bm \Sigma^{1/2}_j(\Z_{js} - \bar{\Z}_j) \big\}
\end{align*}
and $\vecc_{ij, T} = \zeros$ for all pairs $(i, j)$ by construction.
% \begin{align*}
% ||A_{ij, T}(u, h)-B_{ij, T}(u, h)||^2 &= \left(\frac{1}{T}\sum_{t=1}^{T-1} C_{ij, t} K_{t, t+1} \right)^T \left(\frac{1}{T}\sum_{t=1}^{T-1} C_{ij, t} K_{t, t+1}\right) \\
% &= \sum_{d=1}^D \left(\frac{1}{T}\sum_{t=1}^{T-1} C^d_{ij, t} K_{t, t+1} \right)^2\\
% &\leq \frac{1}{T^2} \sum_{d=1}^D \left( \sum_{t=1}^{T-1} \big( C^d_{ij, t} \big)^2 \sum_{t=1}^{T-1} \big(K_{t, t+1}\big)^2 \right) \\
% & =\frac{1}{T^2} \sum_{t=1}^{T-1} \sum_{d=1}^D \big( C^d_{ij, t} \big)^2 \sum_{t=1}^{T-1} \big(K_{t, t+1}\big)^2\\
% & =\frac{1}{T^2} \sum_{t=1}^{T-1} ||C_{ij, t}||^2 \sum_{t=1}^{T-1} \big(K_{t, t+1}\big)^2 \\
% &\leq  \max_{1\leq t\leq T} ||C_{ij, t}||^2 \frac{1}{T}\sum_{t=1}^{T-1} \big(K_{t, t+1}\big)^2
% \end{align*}
From this, it follows that 
{\begin{footnotesize}
\begin{equation} \label{eqA:norm:a-b}
\max_{1\le i < j \le N} \max_{(u,h) \in \mathcal{G}_T} \big|\big|\veca_{ij, T}(u, h)-\vecb_{ij, T}(u, h) \big|\big|_{\infty} \le \max_{(u,h) \in \mathcal{G}_T} W_T(u, h) \max_{1\le i < j \le N}\max_{1 \le t \le T} ||\vecc_{ij, t}||_{\infty}
\end{equation}
\end{footnotesize}}
with $$W_T(u,h) = \frac{1}{\sqrt{Th}}\sum_{t=1}^{T-1}\big|K_{t, u, h} - K_{t + 1, u, h}\big|.$$
Due to the Lipschitz continuity of the kernel function $K(\cdot)$, we have that 
\begin{equation} \label{eqA:kernel_sum}
\max_{(u,h) \in \mathcal{G}_T} W_T(u,h) = O\left( \frac{1}{\sqrt{T (h^{\min}_T)^3}}\right).
\end{equation}
Using the triangular inequality yields that
\begin{align*}
\max_{1 \le t \le T} ||\vecc_{ij, t}||_{\infty} 
 & \le \max_{1 \le t \le T} \Big|\Big| \sum\limits_{s=1}^t \widetilde{\vv}_{is} - \bm \Sigma^{1/2}_i\sum\limits_{s=1}^t \Z_{is} \Big|\Big|_{\infty} + \max_{1 \le t \le T} \Big|\Big| t (\bar{\widetilde{\vv}}_{i} - \bm \Sigma^{1/2}_i\bar{\Z_i}) \Big|\Big|_{\infty}\\
 & \quad + \max_{1 \le t \le T} \Big|\Big| \sum\limits_{s=1}^t \widetilde{\vv}_{js} -  \bm \Sigma^{1/2}_j\sum\limits_{s=1}^t \Z_{js} \Big|\Big|_{\infty} + \max_{1 \le t \le T} \Big|\Big| t (\bar{\widetilde{\vv}}_{j} -\bm \Sigma^{1/2}_j\bar{\Z_j}) \Big|\Big|_{\infty} \\
 & \le 2 \max_{1 \le t \le T} \Big|\Big| \sum\limits_{s=1}^t \widetilde{\vv}_{is} - \bm \Sigma^{1/2}_i\sum\limits_{s=1}^t \Z_{is} \Big|\Big|_{\infty} + 2 \max_{1 \le t \le T} \Big|\Big| \sum\limits_{s=1}^t \widetilde{\vv}_{js} - \bm \Sigma^{1/2}_j\sum\limits_{s=1}^t \Z_{js} \Big|\Big|_{\infty}\\
 &= 2 \max_{1 \le t \le T} \Big|\Big| \sum\limits_{s=1}^t \widetilde{\vv}_{is} - \bm \Sigma^{1/2}_i IB(t) \Big|\Big|_{\infty} + 2 \max_{1 \le t \le T} \Big|\Big| \sum\limits_{s=1}^t \widetilde{\vv}_{js} - \bm \Sigma^{1/2}_j IB(t) \Big|\Big|_{\infty}
\end{align*}
Applying the strong approximation result \eqref{eq-strongapprox-dep} (which is formulated for the Euclidean norm, but the maximum norm is equivalent to it within the fixed number of dimensions $D$), we can infer that
\begin{equation} \label{eqA:strong_approx:result}
\max_{1\le i < j \le N} \max_{1 \le t \le T} ||\vecc_{ij, t}||_{\infty} = o_p\big(T^{1/q}\big).
\end{equation}
Plugging \eqref{eqA:kernel_sum} and \eqref{eqA:strong_approx:result} into \eqref{eqA:norm:a-b}, we obtain that 
\begin{equation} \label{eqA:norm:a-b:result}
\max_{1\le i < j \le n} \max_{(u,h) \in \mathcal{G}_T} \big|\big| \veca_{ij, T}(u, h)-\vecb_{ij, T}(u, h)\big|\big|_{\infty} = o_p \left( \frac{T^{1/q}}{\sqrt{T(h^{\min}_T)^3}} \right).
\end{equation}

Therefore, coming back to \eqref{eqA:psitilde-phidiamond:1}, we obtain that 
{\footnotesize
\begin{align*}
\big| \widetilde{\Psi}_{T} - \Phi^{\diamond}_{T}   \big| 
 & \le \max_{1\le i < j \le N} \big|\big|\widetilde{\bm \Sigma}^{-1}_{ij}\big|\big|_{1} \max_{1\le i < j \le N} \max_{(u,h) \in \mathcal{G}_T} \big|\big|\veca_{ij, T}(u, h) - \vecb_{ij, T}(u, h)\big|\big|_{\infty} \\
 &= O_p(1) \cdot o_p \left( \frac{T^{1/q}}{\sqrt{T(h^{\min}_T)^3}} \right)  = o_p \left( \frac{T^{1/q}}{\sqrt{T(h^{\min}_T)^3}} \right),
\end{align*}}

where we used the fact that $\max_{1\le i < j \le N} \big|\big|\widetilde{\bm \Sigma}^{-1/2}_{ij} \big|\big|_{1}= O_p(1)$, since $\bm \Sigma_{i}$ is positive definite with bounded from $\infty$ and from $0$ eigenvalues for all $i$ according to Assumption \ref{as:x-err5}.

Hence, we just proved \eqref{eqA:psitilde-phidiamond}, i.e., that 
\begin{align*}
    | \widetilde{\Psi}_{T} - \Phi^{\diamond}_{T}  | =    o_p \left( \frac{T^{1/q}}{\sqrt{T(h^{\min}_T)^3}} \right).
\end{align*}

Next, we consider $\big|\Phi^{\diamond}_{T} - \Phi_{T} \big|$ in \eqref{eqA:phidiamond-phi}. Similarly to \eqref{eqA:psitilde-phidiamond:1}, we have that
\begin{align*}
    \big|\Phi^{\diamond}_{T} - \Phi_{T} \big| &\le \max_{1\le i < j \le N} \max_{(u,h) \in \mathcal{G}_T} \big|S^{\diamond}_{ij,T}(u,h) - S_{ij,T}(u,h)\big|\\
    &\leq \max_{1\le i < j \le N} \max_{(u,h) \in \mathcal{G}_T} \big|\big|(\widetilde{\bm \Sigma}^{-1/2}_{ij} - \bm \Sigma_{ij}^{-1/2})\vecb_{ij, T}(u, h)\big|\big|_{\infty} \\
    & \leq \max_{1\le i < j \le N} \max_{(u,h) \in \mathcal{G}_T} \big|\big|\vecb_{ij, T}(u, h)\big|\big|_{\infty} \max_{1\le i < j \le N} \big|\big|\widetilde{\bm \Sigma}^{-1/2}_{ij} - \bm \Sigma_{ij}^{-1/2}\big|\big|_{1}. 
\end{align*}

First, we want to bound $\mathrm{max}_{1\le i < j \le N} \max_{(u,h) \in \mathcal{G}_T} \big|\big|\vecb_{ij, T}(u, h)\big|\big|_{\infty}$:
\begin{align*}
    &\Bigg[\vecb_{ij, T}(u,h)\Bigg]_d = \Bigg[\frac{1}{\sqrt{Th}}\sum^T_{t=1}\bm \Sigma^{1/2}_i\left(\Z_{it} - \bm \bar{\Z}_{i}\right)K_{t, u,h} - \frac{1}{\sqrt{Th}}\sum^T_{t=1}\bm \Sigma^{1/2}_j\left(\Z_{jt} - \bm \bar{\Z}_{j}\right)K_{t, u,h}\Bigg]_d \\
    &= \Bigg[\frac{1}{\sqrt{Th}}\sum^T_{t=1}\bigg(K_{t, u,h} - \frac{1}{T}\sum_{s=1}^T K_{s, u,h}\bigg)\bm \Sigma^{1/2}_i\Z_{it}\Bigg]_d - \Bigg[\frac{1}{\sqrt{Th}}\sum^T_{t=1}\bigg(K_{t, u,h} - \frac{1}{T}\sum_{s=1}^T K_{s, u,h}\bigg)\bm \Sigma^{1/2}_j\Z_{jt}\Bigg]_d \\
    &\quad\quad\sim N\Bigg(0, \frac{1}{Th} \sum^T_{t=1}\bigg(K_{t, u,h} - \frac{1}{T}\sum_{s=1}^T K_{s, u,h}\bigg)^2 \left[\bm \Sigma_i + \bm \Sigma_j\right]_{dd} \Bigg).
\end{align*}

% We can write $\vecb_{ij, T}(u, h) = \vecb^{(I)}_{ij, T}(u, h) - \vecb^{(II)}_{ij, T}(u, h)$ with
% \begin{align*}
%     \vecb^{(I)}_{ij, T}(u, h) &= \frac{1}{\sqrt{Th}}\sum^T_{t=1}\left(\bm \Sigma^{1/2}_i\Z_{it} - \bm \Sigma^{1/2}_j\Z_{jt}\right)K_{t, u, h} \sim N\bigg(\zeros, \frac{(\bm \Sigma_i + \bm \Sigma_j)k_{1, T}(u,h)}{T h}\bigg)\\
%     \vecb^{(II)}_{ij, T}(u, h) &= \frac{1}{\sqrt{Th}}\sum^T_{t=1}\left(\bm \Sigma^{1/2}_i\bar{\Z}_{i} - \bm \Sigma^{1/2}_j\bar{\Z}_{j}\right)K_{t, u, h} \sim N\bigg(\zeros, \frac{(\bm \Sigma_i + \bm \Sigma_j) k_{2,T}(u,h)}{T^2 h}\bigg),
% \end{align*}
We want to use standard results on maxima of Gaussian random variables, and for that we need a uniform bound on the variance. Since $\left[\bm \Sigma_i + \bm \Sigma_j\right]_{dd} \leq C$ for fixed $N$ and $D$, we want to show that $\frac{1}{Th} \sum^T_{t=1}\left(K_{t, u,h} - \frac{1}{T}\sum_{s=1}^T K_{s, u,h}\right)^2 = \frac{1}{Th} \sum^T_{t=1}K^2_{t, u,h} -\frac{1}{T^2 h}\left(\sum_{t=1}^T K_{t, u,h}\right)^2\leq C$ for all $(u, h)$. We can regard this value as a Riemann sum. Hence,
\begin{align*}
    &\lim_{T \to \infty} \frac{1}{T}\sum_{t=1}^T \frac{1}{h}K^2\left(\frac{t/T - u}{h} \right) -\lim_{T \to \infty}\left( \frac{1}{T}\sum_{t=1}^T \frac{1}{\sqrt{h}}K\left(\frac{t/T - u}{h} \right)\right)^2 \\
    &= \lim_{T \to \infty} \frac{1}{T}\sum_{t=1}^T \frac{1}{h}K^2\left(\frac{t/T - u}{h} \right) - \left(\lim_{T \to \infty} \frac{1}{T}\sum_{t=1}^T \frac{1}{h}K\left(\frac{t/T - u}{h} \right)\right)^2  \\
    &= \int_0^1 \frac{1}{h}K^2\left(\frac{z - u}{h}\right) dz - \left(\int_0^1  \frac{1}{\sqrt{h}}K\left(\frac{z - u}{h}\right) dz \right)^2  \\
%    &=\textcolor{red}{\frac{2}{x_2}\int^1_0 \left(K\left(\frac{z-x_{k1}}{x_{k2}}\right) - \int^1_0 K\left(\frac{w-x_{k1}}{x_{k2}}\right)dw\right)^2dz}\\
    &= \int_{-u/h}^{(1 - u)/h} K^2(y) dy - h\left(\int_{-u/h}^{(1 - u)/h} K(y) dy\right)^2\\
    &\leq \int^{1}_{-1} K^2(y) dy < \infty,
\end{align*}
where we used that $K$ has a compact support $[-1, 1]$ and according to the Assumption \ref{as:kernel}, $\int_{-1}^1 K^2(y)dy < \infty$, and for sufficiently large $T$, $\frac{1}{Th} \sum^T_{t=1}\left(K_{t, u,h} - \frac{1}{T}\sum_{s=1}^T K_{s, u,h}\right)^2 \leq 2\int^{1}_{-1} K^2(y) dy < C$ for all $(u, h)$. Therefore, all the coordinates of $\vecb_{ij, T}(u, h)$ are normally distributed random variables with bounded variance. By Assumption \ref{as:GT_bound} and standard results on maxima of Gaussian random variables, we have that
\begin{equation} \label{eqA:norm:b}
\begin{aligned}
    \max_{1\le i < j \le n} \max_{(u,h) \in \mathcal{G}_T} \big|\big|\vecb_{ij, T}(u, h)\big|\big|_{\infty} = O_p(\sqrt{\log T}).
\end{aligned}
\end{equation}
Hence, taking into account that $||\widetilde{\bm \Sigma}^{-1/2}_{ij} - \bm \Sigma_{ij}^{-1/2}||_{1}=o_p(\rho_T)$ for all $i$ and $j$, we have that 
\begin{align*}
    \big|\Phi^{\diamond}_{T} - \Phi_{T} \big| & \leq \max_{1\le i < j \le N} \max_{(u,h) \in \mathcal{G}_T} \big|\big|\vecb_{ij, T}(u, h)\big|\big|_{\infty} \max_{1\le i < j \le N} \big|\big|\widetilde{\bm \Sigma}^{-1/2}_{ij} - \bm \Sigma_{ij}^{-1/2}\big|\big|_{1} \nonumber\\
    &= O_p(\sqrt{\log T}) \cdot o_p(\rho_T) = o_p(\rho_T \sqrt{\log T}). 
\end{align*}
This completes the proof of \eqref{eqA:phidiamond-phi}.
\end{proof}

\begin{proof}[Proof of Proposition \ref{prop:anticon}]
    The proof is an application of anti-concentration bounds for Gaussian random vectors. We in particular make use of the following anti-concentration inequality from \citeapndx{Nazarov2003}, which can also be found as Lemma A.1 in \citeapndx{Chernozhukov2017}. 
%\pagebreak
\begin{lemmaA}\label{lemma-Nazarov}
Let $\boldsymbol{Z} = (Z_1,\ldots,Z_p)^\top$ be a centered Gaussian random vector in $\reals^p$ such that $\ex[Z_j^2] \ge b$ for all $1 \le j \le p$ and some constant $b > 0$. Then for every $\boldsymbol{z} \in \reals^p$ and $a > 0$,
\[ \pr(\boldsymbol{Z} \le \boldsymbol{z} + a) - \pr(\boldsymbol{Z} \le \boldsymbol{z}) \le C a \sqrt{\log p}, \]  
where the constant $C$ only depends on $b$. 
\end{lemmaA}
To apply this result, we introduce the following notation: We write $x = (u,h)$ and $\mathcal{G}_T = \{x_1,\ldots,x_p\}$, where $p := |\mathcal{G}_T| = O(T^\theta)$ for some large but fixed $\theta > 0$ by our assumptions. For $k = 1,\ldots,p$, $d = 1, \ldots, D$ and $1 \le i < j \le N$, we further let 
\begin{align*}
    Z_{ij, 2k-1, d} = \big[\bm \Sigma^{-1/2}_{ij}\vecb_{ij, T}(x_{k1},x_{k2})\big]_d \quad \text{and} \quad Z_{ij, 2k, d} = - \big[\bm \Sigma^{-1/2}_{ij}\vecb_{ij, T}(x_{k1},x_{k2})\big]_d
\end{align*}
along with $\lambda_{ij,2k-1, d} : = \lambda_{ij,2k-1}= \lambda(x_{k2})$ and $\lambda_{ij,2k, d}: = \lambda_{ij,2k} = \lambda(x_{k2})$, where $x_k = (x_{k1},x_{k2})$. Under our assumptions, it holds that $\ex[Z_{ij,l, d}] = 0$ and $\ex[Z_{ij,l, d}^2] \ge b > 0$ for all $i$, $j$, $l$ and $d$. First, as before, 
\begin{align*}
    &\Bigg[\Sigma^{-1/2}_{ij}\vecb_{ij, T}(x_{k1},x_{k2})\Bigg]_d \sim N\Bigg(0, \frac{2}{Tx_{k2}} \sum^T_{t=1}\left(K_{t, x_{k1},x_{k2}} - \frac{1}{T}\sum_{s=1}^T K_{s, x_{k1},x_{k2}}\right)^2 \Bigg).
\end{align*}

We want to show that $\frac{2}{Tx_{k2}} \sum^T_{t=1}\left(K_{t, x_{k1},x_{k2}} - \frac{1}{T}\sum_{s=1}^T K_{s, x_{k1},x_{k2}}\right)^2 = \frac{2}{Tx_{k2}} \sum^T_{t=1}K^2_{t, x_{k1},x_{k2}} -\frac{2}{T^2x_{k2}}\left(\sum_{t=1}^T K_{t, x_{k1},x_{k2}}\right)^2\geq b$ for all $k$. As before, we regard this value as a Riemann sum: 
\begin{align*}
    &\lim_{T \to \infty} \frac{1}{T}\sum_{t=1}^T \frac{2}{x_{k2}}K^2\left(\frac{t/T - x_{k1}}{x_{k2}} \right) -\lim_{T \to \infty}\left( \frac{1}{T}\sum_{t=1}^T \frac{\sqrt{2}}{\sqrt{x_{k2}}}K\left(\frac{t/T - x_{k1}}{x_{k2}} \right)\right)^2 \\
    &= 2\int_{-x_{k1}/x_{k2}}^{(1 - x_{k1})/x_{k2}} K^2(y) dy - 2x_{k2}\left(\int_{-x_{k1}/x_{k2}}^{(1 - x_{k1})/x_{k2}} K(y) dy\right)^2\\
    &\geq 2 \int^{4 - 4x_{k1}}_{-4x_{k1}} K^2(y) dy -\frac{1}{2}\left(\int_{-\infty}^{\infty} K(y) dy\right)^2 \\
    &\geq \int_{-1}^1 K^2(y)dy -\frac{1}{2} = 2b > 0,
\end{align*}
%\commentsB{The part in red is always positive for a bandwidth $h>0$ and a non-constant kernel function.}
where we used the following facts: $x_{k2} \leq h^{\max}_T \leq 1/4$ as stated in Assumption \ref{as:GT_bound}, $K$ is a symmetric function with a compact support $[-1, 1]$, the interval $[-4x_{k1}, 4- 4x_{k1}]$ fully covers either $[-1, 0]$ or $[0, 1]$ for all $x_{k1} \in [0, 1]$, and according to Assumption \ref{as:kernel}, $\int_{-1}^1 K^2(y)dy -\frac{1}{2} > 0$. Therefore, for sufficiently large $T$, we have $\frac{2}{Tx_{k2}} \sum^T_{t=1}\left(K_{t, x_{k1},x_{k2}} - \frac{1}{T}\sum_{s=1}^T K_{s, x_{k1},x_{k2}}\right)^2 \geq \Big(\int_{-1}^1 K^2(y)dy -\frac{1}{2}\Big)/2 = b > 0$.

We next construct the random vector $\boldsymbol{Z} = ( Z_{ij,l, d} : 1 \le i < j \le N, 1 \le l \le 2p, 1\le d \le D)$ by stacking the variables $Z_{ij, l, d}$ in a certain order (which can be chosen freely) and construct the vector $\boldsymbol{\lambda} = (\lambda_{ij,l,d}: 1 \le i < j \le N, 1 \le l \le 2p, 1\le d \le D)$ in an analogous way. Since the variables $Z_{ij,l, d}$ are normally distributed, $\boldsymbol{Z}$ is a Gaussian random vector of length $(N-1)NpD$.

With this notation at hand, we can express the probability $\pr(\Phi_{T} \le q)$ as follows for each $q \in \reals$: 
\begin{align*} 
\pr(\Phi_{T} \le q) 
 & = \pr \Big( \max_{1\leq i< j \leq n}\max_{1 \le l \le 2p} \big\{ \max_{1\le d\le D} Z_{ij,l, d} - \lambda_{ij,l} \big\} \le q \Big) \\
  & = \pr \Big( \max_{1\leq i< j \leq n}\max_{1 \le l \le 2p}  \max_{1\le d\le D} \big\{Z_{ij,l, d} - \lambda_{ij,l, d} \big\} \le q \Big) \\
 & = \pr \big( Z_{ij,l, d} \le \lambda_{ij,l, d} + q \text{ for all } (i,j,l, d) \Big) 
   = \pr \big( \boldsymbol{Z} \le \boldsymbol{\lambda} + q \big). 
\end{align*}
Consequently,
\begin{align*}
\pr\big( |\Phi_{T} - x| \le \delta_T \big) 
 & = \pr \big( x - \delta_T \le \Phi_{T} \le x + \delta_T \big) \\
 & = \pr \big( \Phi_{T} \le x + \delta_T \big) - \pr \big( \Phi_{T} \le x \big) \\
 & \quad + \pr \big( \Phi_{T} \le x \big) - \pr \big( \Phi_{T} \le x - \delta_T \big) \\
 & = \pr \big( \boldsymbol{Z} \le \boldsymbol{\lambda} + x + \delta_T \big) - \pr \big( \boldsymbol{Z} \le \boldsymbol{\lambda} + x \big) \\
 & \quad + \pr \big( \boldsymbol{Z} \le \boldsymbol{\lambda} + x \big) - \pr \big( \boldsymbol{Z} \le \boldsymbol{\lambda} + x - \delta_T\big) \\
 & \le 2 C \delta_T \sqrt{\log((N-1)NpD)},
\end{align*} 
where the last line is by Lemma \ref{lemma-Nazarov}. With $\delta_T = \frac{T^{1/q}}{\sqrt{T (h^{\mathrm{min}}_T})^3} + \rho_T\sqrt{\log T}$, this immediately implies Proposition \ref{prop:anticon}.
\end{proof}

\begin{proof}[Proof of Proposition \ref{prop:size}]
    To proof the result, we first show that
    \begin{align}\label{eq:size_true_test}
        \pr(\Phi_T \leq q_T(\alpha))= 1-\alpha.
    \end{align}
    We prove \eqref{eq:size_true_test} by contradiction. Suppose \eqref{eq:size_true_test} is false. By definition of $q_T(\alpha)$, it holds that $\pr(\Phi_T \leq q_T(\alpha))\geq 1-\alpha$, and thus, there exists $\xi>0$ such that $\pr(\Phi_T \leq q_T(\alpha))= 1-\alpha + \xi$. Following the proof of Proposition \ref{prop:anticon}, we have that for any $\delta>0$,
    \begin{align*}
        \pr(\Phi_T\leq q_T(\alpha)) - \pr(\Phi_T\leq q_T(\alpha) - \delta) \leq 2C\delta\sqrt{\log((N-1)NpD)}.
    \end{align*}
    Combining this result with $\pr(\Phi_T \leq q_T(\alpha))= 1-\alpha + \xi$ yields
    \begin{align*}
        \pr(\Phi_T\leq q_T(\alpha) - \delta) \geq (1-\alpha) + \xi - 2C\delta\sqrt{\log((N-1)NpD)} > (1-\alpha),
    \end{align*}
    for $\delta>0$ small enough. This contradicts the definition of $q_T(\alpha)$, saying that $q_T(\alpha) \coloneqq \inf_{q\in \reals}\{\pr(\Phi_T \leq q)\geq (1-\alpha)\}$. Therefore, by contradiction, \eqref{eq:size_true_test} holds. The result of Proposition \ref{prop:size} follows directly now by applying Theorem \ref{thm:equiv_distr}.
    \begin{align*}
        \big|\pr(\widehat{\Psi}_T \leq q_T(\alpha)) - (1-\alpha)\big| &\overset{\text{(B.10)}}{=}\big|\pr(\widehat{\Psi}_T \leq q_T(\alpha)) - \pr(\Phi_T\leq q_T(\alpha))\big| \\
        &\leq \sup_{y\in\reals}\big|\pr(\widehat{\Psi}_T \leq y) - \pr(\Phi_T\leq y)\big|\\
        &\overset{\text{Thm. 4.1}}{=} o(1).
    \end{align*}
\end{proof}

\begin{proof}[Proof of Proposition \ref{prop:power}]
    
By assumption, there exists a pair $(i_0,j_0)$ and $(u_0,h_0)\in\mathcal G_T$ with $[u_0-h_0,u_0+h_0]\subset[0,1]$ assume that there exists at least one coordinate $d_0\in\{1,\dots,D\}$  such that
\begin{equation}\label{eqA:local_alt}
\big[\bfbeta_{i_0, T}(w)-\bfbeta_{j_0, T}(w)\big]_d \;\ge\; c_T\sqrt{\frac{\log T}{Th}}
\quad\text{for all } w\in [u_0-h_0, u_0+h_0].
\end{equation} 
where $c_T\to\infty$ as $T\to\infty$. Since
\[
\widehat\Psi_T
\;\ge\;
\hat S_{i_0j_0}(u_0,h_0) - \lambda(h_0),
\]
we obtain
\[
\pr\big(\widehat\Psi_T\le q_T(\alpha)\big)
\;\le\;
\pr\big(\hat S_{i_0j_0}(u_0,h_0)-\lambda(h_0)\le q_T(\alpha)\big).
\]
Thus, it suffices to show that
\[
\pr\big(\hat S_{i_0j_0}(u_0,h_0)-\lambda(h_0)\le q_T(\alpha)\big)=o(1).
\]
For notational simplicity, we write $(i,j,u,h)=(i_0,j_0,u_0,h_0)$ as follows.

Using the representation in Section~3.3,
\[
\hat S_{ij}(u,h) = \Bigg|\Bigg|\widehat{\bm\Sigma}^{-1/2}_{ij}\Bigg(\frac{1}{\sqrt{Th}}\sum^T_{t=1}\X_t\bigg(\X_{t}^\prime\bfbeta_{i, T}(t/T) + \varepsilon_{it} - \X_{t}^\prime\bfbeta_{j, T}(t/T) - \varepsilon_{jt} \bigg)K_{t, u, h}\Bigg) \Bigg|\Bigg|_{\infty}.
\]
Hence, we can write
\[
\hat S_{ij}(u,h)
=
\Big\|
A_T(u,h)+B_T(u,h)
\Big\|_\infty,
\]
with
\[
A_T(u,h)
:=
\widehat{\bm \Sigma}_{ij}^{-1/2}\frac{1}{\sqrt{Th}}
\sum_{t=1}^T
\X_t \X_t^\prime (\bfbeta_{i, T}(t/T)-\bfbeta_{j, T}(t/T))\,K_{t,u,h},
\]
\[
B_T(u,h)
:=
\widehat{\bm \Sigma}_{ij}^{-1/2}\frac{1}{\sqrt{Th}}
\sum_{t=1}^T
\X_t (\varepsilon_{it}-\varepsilon_{jt})\,K_{t,u,h}.
\]
The term $A_T(u,h)$ represents the deterministic ``signal'' induced by $\bfbeta_{i}(w)-\bfbeta_{j}(w)$, and $B_T(u,h)$ is the stochastic ``noise'' due to the errors.

For fixed $(i,j,u,h)$, the random vector
\[
\frac{1}{\sqrt{Th}}
\sum_{t=1}^T
\X_t (\varepsilon_{it}-\varepsilon_{jt})K_{t,u,h}
\]
is a kernel-weighted sum of mean-zero weakly dependent random vectors with finite $q$-th moments, by Assumptions \ref{as:err} - \ref{as:x-err4}. Standard moment bounds for such sums (as used in the proof of Theorem~\ref{thm:equiv_distr}) yield that its Euclidean norm is $O_p(1)$, and hence its $\ell_\infty$-norm is also $O_p(1)$.

The long-run covariance estimators satisfy
\[
\|\widehat{\bm \Sigma}_{ij}^{-1/2}-\bm \Sigma_{ij}^{-1/2}\|_1 = o_p(1),
\]
and Assumption \ref{as:x-err5} implies that the eigenvalues of $\bm \Sigma_{ij}$ are bounded away from $0$ and $\infty$. Thus, multiplication by $\widehat{\bm \Sigma}_{ij}^{-1/2}$ preserves stochastic order, and we obtain
\begin{equation}\label{eq:noise-order}
\|B_T(u,h)\|_\infty = O_p(1).
\end{equation}

Now we look at $A_T(u, h)$. By Assumption \ref{as:kernel}, there exists $c_K>0$ such that for all $w$ with $|w-u|\le h \delta_K$,
\begin{equation}\label{eqA:kernel_bound}
K\Big(\frac{w-u}{h}\Big)\ge c_K.
\end{equation}

Consider the signal term before covariance normalization
\[
C_T(u,h)
:=
\frac{1}{\sqrt{Th}}\sum_{t=1}^T
\X_t\X_t^\prime(\bfbeta_{i, T}(t/T)-\bfbeta_{j, T}(t/T))\,K_{t,u,h}.
\]
Since $K_{t,u,h}$ vanishes outside $[u-h,u+h]$, at most $O(Th)$ summands are nonzero; the factor $1/\sqrt{Th}$ therefore yields an overall scale $\sqrt{Th}$.

By Assumption \ref{as:eig_xx} and standard arguments for kernel-weighted laws of large numbers,
\[
\frac{1}{Th}\sum_{t=1}^T \X_t\X_t^\prime K_{t,u,h}
\;\to\; \Gamma(u,h),
\]
where $\Gamma(u,h)$ is positive definite with eigenvalues bounded away from $0$ and $\infty$.

Using \eqref{eqA:local_alt}-\eqref{eqA:kernel_bound}, the boundedness of $\bfbeta_{i}-\bfbeta_{j}$ (from the smoothness of $\bfbeta_i$ and $\bfbeta_j$), and a Riemann-sum approximation, we obtain
\[
\left\|
\frac{1}{Th}\sum_{t=1}^T \X_t\X_t^\prime(\bfbeta_{i, T}(t/T)-\bfbeta_{j, T}(t/T))K_{t,u,h}
\right\|_\infty
\;\ge\; c_1\,c_T\sqrt{\frac{\log T}{T h}}
\]
for some constant $c_1>0$ and all sufficiently large $T$, with probability tending to one. Hence
\begin{align*}
    \|C_T(u,h)\|_\infty &=
\sqrt{Th}
\left\|
\frac{1}{Th}\sum_{t=1}^T \X_t\X_t^\prime(\bfbeta_{i, T}(t/T)-\bfbeta_{j, T}(t/T))K_{t,u,h}
\right\|_\infty\\
&\ge\;c_1\sqrt{Th}\,c_T\sqrt{\frac{\log T}{T h}} \\
&= c_1 \,c_T \sqrt{\log T}
\end{align*}

with probability tending to one.

As before, multiplication by $\widehat{\bm \Sigma}_{ij}^{-1/2}$ preserves the stochastic order, hence, for some constant $c_2>0$,
\[
\|\widehat{\bm \Sigma}_{ij}^{-1/2}v\|_\infty
\;\ge\;
c_2\|v\|_\infty
\quad\text{for all } v\in\mathbb R^D,
\]
with probability tending to one. Applying this to $v=C_T(u,h)$ and combining with the bound on $||C_T(u, h)||_{\infty}$ yields
\begin{equation}\label{eq:signal-order}
\|A_T(u,h)\|_\infty
=
\|\widehat{\bm \Sigma}_{ij}^{-1/2}C_T(u,h)\|_\infty
\;\ge\;
c_3\,c_T\,\sqrt{\log T}
\end{equation}
with probability tending to $1$ for some constant $c_3>0$.

By the triangle inequality and \eqref{eq:noise-order}, \eqref{eq:signal-order},
\[
\hat S_{ij}(u,h)
=
\|A_T(u,h)+B_T(u,h)\|_\infty
\;\ge\;
\|A_T(u,h)\|_\infty - \|B_T(u,h)\|_\infty
\;\ge\;
c_3 \,c_T\,\sqrt{\log T} - O_p(1).
\]
Thus, for any $\varepsilon>0$, there exists $T_0$ such that for all $T\ge T_0$,
\begin{equation}\label{eq:S-lower}
\pr\Big(\hat S_{ij}(u,h) \ge \tfrac{1}{2}c_3 c_T\sqrt{\log T}\Big)
\;\ge\; 1-\varepsilon.
\end{equation}

The multiscale correction term is $\lambda(h) = \sqrt{2\log{1/(2h)}}$. Note that for some sufficiently large constant $C$ we have
\begin{equation}\label{eqA:power:lambda}
\lambda(h) = \sqrt{2\log\{1/(2h)\}} \le \sqrt{2\log\{1/(2h_{\min, T})\}} \le C \sqrt{\log T}.
\end{equation}

The critical value $q_T(\alpha)$ is the $(1-\alpha)$-quantile of the Gaussian max-type statistic $\Phi_T$ defined in~(10). Each component of $\Phi_T$ is the $\ell_\infty$-norm of a Gaussian vector with variance uniformly bounded and bounded away from zero, and the number of components is of order $|\mathcal G_T|\times N^2\times D = O(T^\theta)$ for some $\theta>0$ by Assumption (A12). Standard Gaussian tail bounds for maxima imply that
\begin{equation}\label{eq:qT-order}
q_T(\alpha) = O\big(\sqrt{\log T}\big).
\end{equation}

Combining \eqref{eqA:power:lambda} and \eqref{eq:qT-order}, there exists $C_4>0$ such that
\begin{equation}\label{eq:qT-plus-lambda}
q_T(\alpha)+\lambda(h) \;\le\; C_4\sqrt{\log T}
\quad\text{for all sufficiently large }T.
\end{equation}

From \eqref{eq:S-lower} and \eqref{eq:qT-plus-lambda}, and since $c_T\to\infty$, we can choose $T$ large enough such that
\[
\tfrac{1}{2}c_3\, c_T > C_4,
\]
which implies
\[
\pr\big(\hat S_{ij}(u,h)-\lambda(h) > q_T(\alpha)\big)\to 1.
\]
Equivalently,
\[
\pr\big(\hat S_{ij}(u,h)-\lambda(h)\le q_T(\alpha)\big) = o(1),
\]
which yields the necessary statement.

\end{proof}

\begin{proof}[Proof of Proposition \ref{prop:fwer}]
    The proof of this result follows the steps in the proof of Proposition 4.5 in \citeapndx{KhismatullinaVogt2022} and therefore omitted for brevity.
\end{proof}

\newpage
\clearpage
\bibliographystyleapndx{ims}
\bibliographyapndx{bib}
\end{spacing}
\endgroup

\end{document}